\documentclass[letterpaper]{JHEP3} 
\usepackage{epsfig,array,cite,latexsym,amsmath,oldgerm}
\preprint{RIKEN-TH-99}
\usepackage{amsmath,amssymb}
\usepackage[all]{xy}
\makeatletter
 
  \@addtoreset{equation}{section}
 \makeatother

\newcommand\half{{\textstyle{\frac{1}{2}}}}

\newcommand{\be}{\begin{equation}}
\newcommand{\ee}{\end{equation}}
\newcommand{\bfs}{\mathbf{S}}
\newcommand{\phib}{\ensuremath{\overline{\phi}}}
\newcommand{\cF}{\ensuremath{{\cal F}}}
\newcommand{\bfz}{\mathbf{Z}}
\newcommand{\bea}{\begin{eqnarray}}
\newcommand{\eea}{\end{eqnarray}}
\newcommand{\beann}{\begin{eqnarray*}}
\newcommand{\eeann}{\end{eqnarray*}}
\newcommand{\nn}{\nonumber}
\newcommand{\ba}{\begin{array}}
\newcommand{\ea}{\end{array}}
\newcommand{\xh}{\mathbf{\hat{e}}_1}
\newcommand{\yh}{\mathbf{\hat{e}}_2}

\newcommand{\ih}{\mathbf{\hat{e}}_i}
\newcommand{\jh}{\mathbf{\hat{e}}_j}

\newcommand{\LR}{\Leftrightarrow}
\newcommand{\mybar}[1]%
        {\kern 0.6pt\overline{\kern -0.6pt#1\kern -0.6pt}\kern 0.6pt}

\newcommand{\Tr}{{\rm Tr}\,}

\newcommand{\Z}{\mathbb{Z}}
\newcommand{\R}{\mathbb{R}}

\newcommand{\nv}{{\bf n}}

\newcommand{\iv}{{\bf i}}

\title{Relationship
between 
various supersymmetric lattice models}
\author{Tomohisa Takimi
\\
Theoretical Physics Laboratory
The Institute of Physical and Chemical Research (RIKEN)
2-1 Hirosawa, Wako
Saitama 351-0198, JAPAN \\
Email:\email{ttakimi@riken.jp}}
\abstract{
We comment on the relationships between several supersymmetric lattice models; 
the ``orbifold lattice theory'' by Cohen-Kaplan-Katz-Unsal (CKKU),
lattice regularization of the topological field theory by Sugino and the
``geometrical approach'' by Catterall. We point out that these three models
have close relationships; the ${\cal N} =(2,2)$ model by 
Catterall~\cite{Catterall:2004np}
and the two-dimensional ${\cal N} = (2,2)$ lattice
theory being similar to Sugino's construction~\cite{Sugino:2003yb} 
can be derived by
appropriate truncation of fields in the two-dimensional ${\cal N} = (4,4)$
orbifold lattice theory by CKKU~\cite{Cohen:2003qw}. 
Catterall's ${\cal N} = (2,2)$ description possesses extra degrees of
freedom compared to the target ${\cal N} = (2,2)$ theory. If we remove
those extra degrees of freedom in a way keeping supersymmetry on the lattice,
Catterall's description reduces to a model of the Sugino type.}
\begin{document}
\tableofcontents
\section{Introduction}
Recently, several lattice gauge theories which preserve
partial supersymmetry on the lattice are
proposed~\cite{Kaplan:2002wv,Cohen:2003xe,Cohen:2003qw,Kaplan:2005t,
Endres:2006ic,
Damgaard:2007be,
Sugino:2003yb,Sugino:2004qd,Sugino:2004gz,
Sugino:2004uv,Sugino:2006uf,Catterall:2004np,Catterall:2005fd,
Catterall:2006jw}.\footnote{There are other several supersymmetric lattice
models which are not treated in this
paper~\cite{Catterall:2001fr,
Unsal:2004cf,
Giedt:2003ve,
Suzuki:2005dx, 
Feo:2002yi,
Montvay:2001kw,
Kaplan:1999jn,
Montvay:2001aj,
DAdda:2005zk}.}
The main purpose in these models is to solve the fine-tuning problem in
lattice regularizations of supersymmetric gauge theories.
The models utilize the
topological twisting~\cite{Witten:1988ze,Unsal:2006qp} to pick up the
subset of superalgebra which does not include the infinitesimal translation.
In this way, the partial supersymmetry can be preserved on the lattice which
explicitly breaks the infinitesimal translational invariance.

There are several types of the model: The series of models proposed by
Cohen-Kaplan-Katz-Unsal-Endres~\cite{Kaplan:2002wv,
Cohen:2003xe,Cohen:2003qw,Kaplan:2005t,Endres:2006ic} are ``orbifold lattices''
which are constructed from reduced supersymmetric matrix models by the 
orbifold projection~\cite{Douglas:1996sw} and the
deconstruction~\cite{Arkani-Hamed:2001ie}.
In their way, the orbifold projection generates the lattice theory with
preserved subset of supersymmetry of the target theory. The deconstruction
dynamically generates space-time by the vacuum expectation value
$\frac{1}{\sqrt{2}a}$ of bosonic
link fields, where $a$~denotes the lattice spacing.

The other approach, proposed by
Sugino~\cite{Sugino:2003yb,Sugino:2004qd,Sugino:2004gz,Sugino:2004uv,
Sugino:2006uf}, are lattice regularizations of the ``topological field theory
action'' which is equivalent to the extended supersymmetric gauge theory.
In his approach, the BRST-like supercharges are preserved on the lattice
because such charges do not generate the infinitesimal translation.

Catterall proposed
models~\cite{Catterall:2004np,Catterall:2005fd,Catterall:2006jw} which are
based on the Kahler-Dirac formalism and the lattice analogue of
differential forms~\cite{Aratyn:1984bd}. In his models, the 1-form and 2-form
fields have to be complex because they are in the bi-fundamental representation
of the lattice gauge group and the hermiticity cannot be maintained under gauge
transformations. Since the counterparts of these 1- and 2-form fields in the
target theory are hermitian, Catterall's models have extra degrees of freedom
which we have to discard in the path-integral. If one performs such
truncation in a naive way, supersymmetry on the lattice would be broken.

Seemingly, these three types of model are quite different. There exist,
however, close relationships between them. 
We will clarify such relationships in this paper.
This investigation of the relationships would be very useful
to develop the lattice formulations of
supersymmetric theories.
First, in section~2, we show that Catterall's ${\cal N} = (2,2)$ 
action~\cite{Catterall:2004np} can be embedded in CKKU's
${\cal N} = (4,4)$ action~\cite{Cohen:2003qw} 
under suitable field truncation. Then, in section~3,
we explain the relationship between Catterall's ``complexified''
${\cal N} =(2,2)$ lattice theory and Sugino's theory of
ref.~\cite{Sugino:2003yb}. For Catterall's model to contain the correct numbers
of degrees of freedom compared to 
the target ${\cal N} = (2,2)$ theory, we have to
truncate extra degrees of freedom. If we perform the truncation in a way
keeping the supersymmetry on the lattice, Catterall's model becomes 
the ${\cal N} =(2,2)$ model being similar to Sugino's model. 
Finally, in section~4, we explain that the ${\cal N} = (2,2)$ 
supersymmetric lattice model of the Sugino type 
can be directly derived from CKKU's
${\cal N} = (4,4)$ lattice theory by restricting fields.
We also explain that the derivation discards the quantum fluctuations 
of scalar zero modes around the
vacuum expectation value $\frac{1}{\sqrt{2}a}$. 
In section 5, we also give an continuum analogue 
of the truncation of degrees of freedom.
By this truncation, we can obtain the continuum ${\cal N} = (2,2)$
super Yang-Mills theory from the continuum ${\cal N} = (4,4)$
super Yang-Mills theory.
Section 6 is
devoted to conclusion and discussion. 

\section{Relationship between the ${\cal N} = (4,4)$ CKKU model and
the ${\cal N} = (2,2)$ Catterall model}
Here we explicitly show that Catterall's ${\cal N} = (2,2)$ lattice model
can be obtained by truncating certain fields in the ${\cal N} = (4,4)$ CKKU
model.
%
%
%
%
\subsection{${\cal Q}$-exact form of  the ${\cal N} = (4,4)$ CKKU model}
We explain the ${\cal N} = (4,4) $ supersymmetric lattice theory proposed by
CKKU in ref.~\cite{Cohen:2003qw} very briefly. 
Here we neglect the soft supersymmetry breaking mass term in their 
theory.\footnote{In CKKU's models, there are flat directions in the scalar
potential allowing large fluctuations around the vacuum expectation
value~$\frac{1}{\sqrt{2}a}$. To stabilize the lattice structure, CKKU
introduced the soft SUSY breaking mass terms. In this section, we will
investigate their model without such mass terms.}
The action of the theory
(eq.~(3.14) of ref.~\cite{Cohen:2003qw}) is
\begin{equation}\begin{aligned}
S= \sum_{\nv}\Tr &\left[\int d\theta d\mybar\theta\,  \left(
    \frac{1}{2}\mybar{\bold\Upsilon}_{\nv}{\bold\Upsilon}_{\nv} 
+\frac{1}{\sqrt{2}}{\bfs}_{\nv}( {\bfz}_{i,\nv}{{\mybar\bfz}}_{i,\nv} - {{\mybar\bfz}}_{i,\nv-
    \ih} {\bfz}_{i,\nv-\ih})
-\frac{1}{2}\mybar{\bold\Xi}_{\nv}
{\bold \Xi}_{\nv}  \right) \right.\\ 
&\quad +  \left.
\int d\theta\,\Bigl(
\epsilon_{ij}\, {\bold \Xi}_{\nv }\,{\bfz}_{i,\nv}{\bfz}_{j,\nv
    +\ih} 
\Bigr)  
- \int d\mybar\theta\,\Bigl( 
\epsilon_{ij}\,\mybar{\bold\Xi}_{\nv
}\,\mybar {\bfz}_{i,\nv+\jh}\mybar {\bfz}_{j,\nv} 
\Bigr) 
\right],
\end{aligned}\label{lact2}\end{equation}
where the sum over site $\nv=\{n_1,n_2\}$ is taken in the interval $n_{1,2}\in
[1,N]$ and $\xh$ and~$\yh$ being unit vectors in $n_1$ and~$n_2$ directions,
respectively. The superfields are defined by
\begin{equation}
\begin{aligned}
{\bfz_{i,\nv}} &= z_{i,\nv} +\sqrt{2}\,\theta \psi_{i,\nv} -\sqrt{2}\, \theta\mybar\theta
(\mybar z_{3,\nv} z_{i,\nv}- z_{i,\nv} \mybar z_{3,\nv + \ih})\ ,\\
\mybar {\bfz}_{i,\nv} &= \mybar z_{i,\nv}
+\sqrt{2}\,\mybar\theta \epsilon_{ij}\xi_{j,\nv} +\sqrt{2}\, \theta\mybar\theta
(\mybar z_{3,\nv+\ih}\mybar z_{i,\nv} -\mybar z_{i,\nv}\mybar z_{3,\nv})  \ , \\
%
{\bold \Xi_\nv} &= \xi_{3,\nv} +\sqrt{2}\, \theta
\tilde{G}_\nv
-\sqrt{2}\, \theta \mybar\theta
(\mybar z_{3,\nv + \xh + \yh}\xi_{3,\nv} -\mybar z_{3,\nv}\xi_{3,\nv}
)\ ,\\
\mybar{\bold \Xi}_{\nv} &= \chi_\nv -\sqrt{2}\,\mybar\theta
\tilde{\bar{G}}_\nv
 +\sqrt{2}\, \theta \mybar\theta
(\mybar z_{3,\nv}\chi_\nv - \chi_\nv\mybar z_{\nv+\xh+\yh}) \ ,\\
{\bfs_\nv} &= z_{3,\nv} + \sqrt{2}\,\theta \psi_{3,\nv} + \sqrt{2}\, \mybar\theta \lambda_\nv
+\sqrt{2}\,\theta \mybar\theta 
i\tilde{d}_{\nv}
\ ,\\
{\bold \Upsilon}_\nv &=  \lambda_\nv - 
\theta \Bigl(
i \tilde{d}_\nv+[\mybar{z}_{3,\nv}, z_{3,\nv}] +\Bigr) 
-\sqrt2 \theta \mybar{\theta}[\mybar{z}_{3,\nv}, \lambda_\nv]\ ,\\
\mybar {\bold\Upsilon}_\nv& = \psi_{3,\nv} + \mybar{\theta} \Bigl(
i \tilde{d}_\nv -[\mybar{z}_{3,\nv}, z_{3,\nv}] \Bigr)
+ \sqrt2 \theta \mybar{\theta}[\mybar{z}_{3,\nv}, \psi_{3,\nv}]\ ,
\end{aligned}
\label{sfields2d}\end{equation}
with 
\be
\begin{split}
\tilde{\bar{G}}_\nv =
\mybar G_\nv-\sqrt{2}\,\epsilon_{ij}\, z_{i,\nv}  z_{j,\nv+\ih},
\\
\tilde{G}_\nv =
G_\nv-\sqrt{2}\,\epsilon_{ij}\,\mybar z_{i,\nv+\jh}\mybar
z_{j,\nv},
\\
\tilde{d}_\nv = d_\nv 
-i(\mybar z_{i,\nv-\ih}  z_{i,\nv-\ih} - z_{i,\nv}\mybar z_{i,\nv}).
\end{split}
\ee
In these expressions, $\theta$, $\bar{\theta}$ are one-component Grassmann
super coordinates. All variables are $M\times M$ matrices satisfying periodic
boundary conditions on the lattice, and there is an independent $U(M)$ symmetry 
associated with each site which becomes the $U(M)$ gauge symmetry of the
continuum theory. The indices $i$, $j$ run over $1$ and~$2$ and
all repeated indices are summed. The variables $z_a$ $(a = 1,2,3)$ and  
$\mybar z_a$ refer to complex bosonic variables and their conjugates, while
$\lambda$, $\chi$, $\psi_a$ and $\xi_a$ refer to one-component Grassmann
variables. Here $d_\nv$, $G_\nv$, $\bar{G}_\nv$ are auxiliary fields originally
introduced in ref.\cite{Cohen:2003qw}. These auxiliary fields are
integrated out yielding $d_\nv = G_\nv = \bar{G}_\nv = 0 $.

The supersymmetry on the lattice can be read off from eq.~(\ref{sfields2d}).
It is
\be
\begin{split}
\delta z_{i,\nv} &= i\sqrt{2}\,\eta\,\psi_{i,\nv}, 
\\
\delta\mybar z_{i,\nv}
&=i\epsilon_{ij}\,\sqrt{2}\,\mybar\eta\,\xi_{j,\nv},\\
\delta\psi_{i,\nv} &= 2i\mybar\eta\,
[z_{i,\nv}\mybar z_{3,\nv+\ih} -\mybar z_{3,\nv}z_{i,\nv}],  
\\
\delta\xi_{i,\nv} &= -2i\epsilon_{ij}\,\eta\,
[\mybar z_{j,\nv}\mybar z_{3,\nv}- \mybar z_{3,\nv+\jh}
\mybar z_{j,\nv}],\\  
\delta z_{3,\nv} &= i\sqrt{2}\,
(\eta\,\psi_{3,\nv} +\mybar \eta\, \lambda_{\nv}), \\
 \delta\mybar z_{3,\nv}
&= 0,\\
\delta\psi_{3,\nv} &= i\mybar\eta\,
\left([\mybar z_{i,\nv-\ih}z_{i,\nv-\ih}-z_{i,\nv}\mybar z_{i,\nv}]
 - [\mybar z_{3,\nv},\,z_{3,\nv}]
+i d_\nv\right),
\\
\delta\lambda_{\nv} &= -i\eta\,
\left([\mybar z_{i,\nv-\ih}z_{i,\nv-\ih}
- z_{i,\nv}\mybar z_{i,\nv}
]
+ [\mybar
  z_{3,\nv},\,z_{3,\nv}] + i d_\nv \right), \\  
\delta\chi_\nv &=i \mybar\eta\,[2(z_{1,\nv}z_{2,\nv+\xh}-
z_{2,\nv}z_{1,\nv+\yh}) -\sqrt{2}\,\, \mybar G_\nv],
\\
  \delta\xi_{3,\nv} &= -i\eta\,
[2(\mybar z_{1,\nv+\yh}\mybar z_{2,\nv}-\mybar z_{2,\nv+\xh}\mybar z_{1,\nv})
-\sqrt{2} G_\nv],
\end{split}
\label{supercharge}
\ee
and 
\be
\begin{split}
\delta \mybar G_\nv
&= 2i\eta\,\left( \epsilon_{ij} \,(z_{i, \nv} \psi_{j, \nv + \ih}
-\psi_{j, \nv}  z_{i, \nv+\jh} )
+(\mybar z_{3,\nv} \chi_\nv -\chi_\nv \mybar z_{3,\nv+ \xh+\yh})
\right),
\\
\delta G_\nv
&= -2i\mybar \eta\, 
\left(
\sum_{i,j,\,\text{with}\, i\ne j}(\mybar z_{i, \nv+\jh} \xi_{i, \nv}
-\xi_{i, \nv+\ih}  \mybar z_{i, \nv} )
+(\mybar z_{3,\nv+\xh+\yh} \xi_{3,\nv} -\xi_{3,\nv} \mybar z_{3,\nv})
\right),
\\
\delta d_\nv 
&= -\sqrt{2} \eta
(\mybar z_{i, \nv-\ih}\psi_{i,\nv-\ih}
-\psi_{i,\nv}\mybar z_{i, \nv} 
+ [\mybar z_{3,\nv}, \psi_{3,\nv}])
\\
& \quad\quad +\sqrt{2} \mybar \eta
(\epsilon_{ij} (z_{i, \nv}\xi_{j,\nv}
-\xi_{j,\nv-\ih}z_{i, \nv-\ih} )
+ [\mybar z_{3,\nv}, \lambda_{\nv}]).
\end{split}
\label{aux-trans}
\ee
We may express the supersymmetry transformation by using two supercharges
\be
\delta = i\eta Q + i\mybar \eta \mybar Q.
\label{q2susy}
\ee
These charges $Q$, $\mybar Q$ can be realized in terms of 
independent Grassmann coordinates $\theta$ and $\mybar \theta$ as
\begin{equation}
Q =\frac{\partial\  }{\partial \theta} 
+ \sqrt{2}\, \mybar\theta[\mybar z_3,\,\cdot \ ]^{\ast} \ ,\qquad
\mybar Q =\frac{\partial\  }{\partial \mybar\theta} +   
\sqrt{2}\, \theta[\mybar z_3,\,\cdot \ ]^{\ast} \ ,
\label{repre-alg}
\end{equation}
where the operation~$[\mybar z_3,\,\cdot \ ]^{\ast}$ represents
the lattice gauge transformation with the parameter $\mybar z_3$.
This operation $[\mybar z_3,\,\cdot \ ]^{\ast}$ acts on generic fields
$P_\nv$ living on the links as
\be
[\mybar z_3,\,P_\nv ]^{\ast}
\equiv
\mybar z_{3,\nv} P_\nv - P_\nv \mybar z_{3,\nv + r_i\ih},
\ee
where we have assumed that the link under consideration connects two sites
$\nv$ and~$\nv+r_i \ih$. This rule is applied to
$z_{i,\nv}$, $\psi_{i,\nv}$, $\chi_\nv$ and~$\mybar G_\nv$. Similarly, for the
anti-oriented link fields $\mybar P_\nv$, such as
$\mybar z_{i,\nv}$, $\xi_{i,\nv}$, $\xi_{3,\nv}$ and~$G_\nv$,
\be
[\mybar z_3,\,\mybar P_\nv ]^{\ast}
\equiv
\mybar z_{3,\nv+ r_i\ih} \mybar P_\nv - \mybar P_\nv \mybar z_{3,\nv}.
\ee
For site fields $P'_\nv$, which are $z_{3,\nv}$, $\lambda_\nv$,
$\psi_{3,\nv}$ and~$d_\nv$, the operation is simply the commutator 
$[\mybar z_{3,\nv}, P'_\nv]^{\ast} \equiv
[\mybar z_{3,\nv}, P'_\nv]$.
Auxiliary fields $G_\nv$, $\mybar G_\nv$, $d_\nv$
and their transformation laws~(\ref{aux-trans})
are introduced to make the algebra,\footnote{Be careful that there is the minus
sign in eq.~(\ref{CKKU-algebra}) which 
cannot be appear from the anti-commutation
relation in the representation of super coordinates eq.~(\ref{repre-alg}).
This difference comes from the fact 
that the left operation of supersymmetry group 
corresponds to the right motion in parameter space as described in 
the textbook written by Wess-Bagger~\cite{Wess-Bagger}.}
$Q^2=\mybar Q^2 = 0$ and
\be
\{ Q, \mybar Q \} = -2\sqrt{2} [\mybar z_3, \cdot ]^{\ast},
\label{CKKU-algebra}
\ee
to hold off-shell.

We define the BRST-like charge ${\cal Q}$ by
\be
{\cal Q} = \frac{1}{\sqrt{2}} 
\left( Q - \mybar Q \right),
\ee
which induces
\be
\begin{array}{ll}
{\cal Q}z_{i,\nv}
=\psi_{i,\nv}, & {\cal Q}\psi_{i,\nv} =\sqrt{2}
(\mybar z_{3,\nv} z_{i,\nv} - z_{i,\nv}\mybar z_{3,\nv+\ih}),\\
{\cal Q}\mybar z_{i,\nv}
= -\epsilon_{ij}\xi_{j,\nv}, & {\cal Q}\xi_{i,\nv} =\sqrt{2}
\epsilon_{ij}
(\mybar z_{3,\nv+\jh} \mybar z_{j,\nv} 
- \mybar z_{j,\nv}\mybar z_{3,\nv}),\\
{\cal Q}z_{3,\nv}
=\psi_{3,\nv} - \lambda_\nv, & 
{\cal Q}(\psi_{3,\nv} - \lambda_\nv)
=\sqrt{2}
[\mybar z_{3,\nv},z_\nv],\\
{\cal Q}\tilde{d}_\nv =
i[\mybar z_{3,\nv},\psi_{3,\nv} + \lambda_\nv], &
{\cal Q}(\psi_{3,\nv} + \lambda_\nv) = -\sqrt{2} i\tilde{d}_\nv,\\
{\cal Q}\chi_\nv 
= \tilde{\mybar G}_\nv,
&
{\cal Q}\tilde{\mybar G}_\nv=
\sqrt{2}(\mybar z_{3,\nv}\chi_\nv-\chi_\nv \mybar z_{3,\nv+\xh+\yh}),
\\
{\cal Q}\xi_{3,\nv} 
= \tilde{G}_\nv,
&
{\cal Q}\tilde{G}_\nv=
\sqrt{2}(\mybar z_{3,\nv+\xh+\yh}\xi_{3,\nv}-
\xi_{3,\nv}\mybar z_{3,\nv}),
\\
{\cal Q}\mybar z_{3,\nv}=0. & 
\end{array}\label{orb-BRST}
\ee
This charge ${\cal Q}$ also satisfies 
\be
{\cal Q}^2 = \sqrt{2}[\mybar z_3 , \cdot]^{\ast},
\ee
where the right hand side 
is the gauge transformation with the parameter $\mybar z_3$.

Now, for our purpose, it is crucial to rewrite the action~(\ref{lact2})
in a ${\cal Q}$ exact form. 
Then it can be confirmed that the action is ${\cal Q}$ exact
\be
S = \frac{1}{2g^2} {\cal Q} \,\Xi,
\ee
\be
\begin{split}
\Xi
&= \sum_{\nv}
\Tr\Bigl[\frac{1}{\sqrt{2}}(\psi_{3,\nv} - \lambda_\nv)
[\mybar z_{3, {\bf n}},z_{3, {\bf n}}]
+\frac{1}{\sqrt{2}}(\psi_{3,\nv} + \lambda_\nv)
[i\tilde{d}_\nv -2
(\mybar z_{i,\nv -\ih}z_{i,\nv -\ih} -
z_{i,\nv}\mybar z_{i,\nv})]
\\
&\quad \qquad \qquad
+\xi_{3,\nv}(
\tilde{\mybar G}_\nv + 2\sqrt{2}\epsilon_{ij}
z_{i,\nv}z_{j,\nv+\ih})
+\chi_\nv
(\tilde{G}_\nv + 2\sqrt{2}\epsilon_{ij}
\mybar z_{i,\nv+\jh}\mybar z_{j,\nv})
\\
&\qquad\quad+\sqrt{2}
\psi_{i,\nv}(\mybar z_{i,\nv}z_{3,\nv}
-z_{3,\nv+\ih}\mybar z_{i,\nv})
+\sqrt{2}
\epsilon_{ij}\xi_{i,\nv-\jh}(z_{j,\nv-\jh}z_{3,\nv}
-z_{3,\nv-\jh}z_{j,\nv-\jh})
\Bigr].
\end{split}
\label{orb-act-2}
\ee
We will use this form to clarify the relationships.

%
%
%
\subsection{The ${\cal N} = (2,2)$ Catterall model}
\label{Sec:cat-rev0}
Catterall's ${\cal N} = (2,2)$ supersymmetric lattice gauge
theory~\cite{Catterall:2004np} is based on the fact that the
${\cal N} = (2,2)$ supersymmetric Yang-Mills theory can be
regarded equivalently as a topological field theory~\cite{Witten:1988ze}.
The continuum action is thus expressed by an exact form by using
a nilpotent supercharge $Q$
\be
S=\beta Q{\rm Tr}\int
d^2x\left(
\frac{1}{4}\eta[\phi,\phib]-2i\chi_{12}F_{12}+\chi_{12}B_{12}+\psi_\mu D_\mu \phib\right)
\label{gfermion}
\ee
where $\phi$, $\phib$ are bosonic scalar fields, 
$B_{12}$ is a bosonic anti-symmetric two tensor field,
$F_{12}$ is a field strength of vector gauge fields $A_{\mu}$.
$\eta$, $\psi_\mu$, $\chi_{12}$ are fermion fields with one component spinor
index.
$\eta$ is regarded as a scalar and $\psi_\mu$ are vectors
and $\chi_{12}$ is an antisymmetric two-tensor
under twisted rotational symmetry. 
Parameter $\beta$ represent the inverse of the square of 
gauge coupling.
Here,
all fields are taken in the adjoint representation
$C = \sum_a C^a T^a$ where $T^a$ are
\textit{anti-hermitian generators} in the gauge 
group and $C^a$ are real.
The gauge symmetry is unitary $U(M)$.
$D_\mu$ is a covariant derivative with the adjoint representation
using the anti-hermitian matrices $A_\mu$.
Indices $\mu$ run from $1$ to $2$ which represent the directions in 
two dimensional Euclidean space.

\subsubsection{Catterall's lattice action}
In constructing the lattice action, Catterall utilizes the Kahler-Dirac
formalism and the lattice analogue of differential forms.
He applied the criterion such that each scalars, vectors and antisymmetric 
two-tensors should be put on sites, links, and the plaquettes, respectively,
on the lattice. Therefore,
scalar fields $\phi$, $\phib$ and $\eta$ are put on sites
and vectors $A_{\mu}$, $\psi_{\mu}$ reside on links and
anti-symmetric two tensors $\chi_{12}$, $B_{12}$ reside on plaquettes.
Then Catterall's action is given as\footnote{
We change the notation of Catterall model a little bit.
The difference from the original notation in his 
papers~\cite{Catterall:2004np,Catterall:2006jw} 
is as follows:
In the action~(\ref{caterall-act}),
we change the parameter
$\beta$ as
$-\beta$. 
To take the 
continuum limit~(\ref{gfermion})
consistent with the anti-hermitian condition
$\eta^\dag = -\eta$ imposed later on,
this change is required.
We also change the notation 
$\chi_{12}^\dag\cF_{12}$, $\chi_{12}\cF_{12}^\dag$ to
$-i\chi_{12}^\dag\cF_{12}$, $-i\chi_{12}\cF_{12}^\dag$. 
By this change, the kinetic term of gauge field
can be taken as positive definite
$\cF_{12}\cF^\dag_{12}$ after the integration of 
auxiliary fields $B_{12}B_{12}^\dag$.
If we do not change the notation,
kinetic term of the gauge fields becomes
$-\cF_{12}\cF^\dag_{12}$.
For the same reason, we also change the $2\chi_{12}\cF_{12}$ to
$-i2\chi_{12}\cF_{12}$ in the target action~(\ref{gfermion}).}
\begin{eqnarray}
S_L&=&-\beta Q{\rm Tr}
\sum_{x}\left(\frac{1}{4}\eta^\dagger(x)[\phi(x),\bar{\phi}(x)]-i
\chi^\dagger_{12}(x){\cal F}_{12}(x)-i\chi_{12}(x){\cal F}_{12}(x)^\dagger\right.\nonumber\\
&+&
\biggl( \frac{1}{2}\chi^\dagger_{12}(x)B_{12}(x)+\frac{1}{2}\chi_{12}(x)B^\dagger_{12}(x) \nn \\
&&
+\frac{1}{2}\psi^\dagger_\mu(x)D^+_\mu\bar{\phi}(x)+\frac{1}{2}\psi_\mu(x)(D^+_\mu\bar{\phi}(x))^\dagger \biggr)
\label{caterall-act}
\end{eqnarray}
where $U_{1,2}$ are bosonic link variables defined as $U_{\mu} = e^{A_\mu}$,
and ${\cal F}_{12}$ are field strength of gauge fields
defined as
\be
\cF_{12}(x)=D^+_1 U_2(x)=U_1(x)U_2(x+1)-U_2(x)U_1(x+2)
\ee
whose continuum limit is $F_{12}(x)$.
$D^+_\mu$ is covariant version of forward difference 
acting on a scalar field $f(x)$ and a vector field $f_{\mu}(x)$
as~\cite{Aratyn:1984bd}
\begin{eqnarray}
D^+_\mu f(x) &=& U_\mu(x)f(x+\mu)-f(x)U_\mu(x),\nonumber\\
D^+_\mu f_\nu(x)&=& U_\mu(x)f_\nu(x+\mu)-f_\nu(x)U_\mu(x+\nu).
\end{eqnarray}
Note that, compared to the target theory~(\ref{gfermion}), several new fields 
$\eta^\dag$, $\phi^\dag$, $\phib^\dag$, $\psi^{\dag}_\mu$, $\chi^\dag$
and~$B^\dag_{12}$,
appear in his action. These conjugate fields transform as complex conjugate 
of original fields $\eta$, $\phi$, $\phib$, $\psi_\mu$, $\chi$ and~$B_{12}$
under gauge transformation.
Such conjugate fields are required to preserve the lattice gauge symmetry
and naturally appear in the Kahler-Dirac formulation as described in
section 3 of ref.~\cite{Catterall:2004np}.

His $Q$ transformation is defined by
\be
\begin{array}{ll}
QU_\mu=\psi_\mu &
QU^\dagger_\mu=\psi^\dagger_\mu,\nonumber\\
Q\psi_\mu=-D^+_\mu\phi, &
Q\psi^\dagger_\mu=-(D^+_\mu\phi)^\dagger,\nonumber\\
Q\chi_{12}=B_{12}, &
Q\chi^\dagger_{12}=B^\dagger_{12},\nonumber\\
QB_{12}=[\phi,\chi_{12}]^{(12)}, &
QB^\dagger_{12}=\left([\phi,\chi_{12}]^{(12)}\right)^\dagger,\nonumber\\
Q\phib=\eta, &
Q\phib^\dag=\eta^\dag,\nonumber\\
Q\eta=[\phi,\phib], &
Q\eta^\dag = ([\phi,\phib])^\dag,  \nn\\
Q\phi=0, &
\end{array}
\label{cat-q}
\ee
where the superscript notation indicates a {\it shifted} commutator
\be
[\phi,\chi_{\mu\nu}]^{(\mu\nu)}=\phi(x)\chi_{\mu\nu}(x)-
\chi_{\mu\nu}(x)\phi(x+\mu+\nu).\ee
Note that the $Q$-transformation laws satisfy following property 
\be
Q^2 = (\text{gauge transformation with the parameter}\,\, \phi).
\label{nil-gauge}
\ee 

\subsubsection{Extra degrees of freedom in Catterall's theory}
\label{Sec:cat-act}
In the lattice action~(\ref{caterall-act}),
there are extra degrees of freedom 
which the target theory does not have.
Variables $\phi$, $\phib$, $\eta$, $\psi_\mu$, $\chi,B_{12}$, $A_\mu$ 
on the lattice 
are defined with
general complex matrices $C= \sum_a (C^a+ i D^a) T^a$, 
where $C^a,D^a$ are real,
while the variables in the target theory
are defined with anti-hermitian matrices $C = \sum C^a T^a$.
This is because 
the vector and tensor fields reside
on the links and plaquettes, 
which are bi-fundamental representation under the lattice
gauge group.
The gauge transformation laws of generic vector fields $f_{\mu}(x)$ and 
anti-symmetric two tensors $f_{\mu\nu}(x)$ are assumed to be
\bea
f_{\mu}(x) \to V(x) f_{\mu}(x)V(x+\mu)^\dag ,
\nn \\
f_{\mu\nu}(x) \to V(x) f_{\mu\nu}(x)V(x+\mu+\nu)^\dag,
\eea
where $V(x)$, $V(x+\mu)$ and $V(x+\mu+\nu)$ are independent unitary matrices.
Anti-hermiticity of the bi-fundamental variables 
cannot be maintained under the gauge transformation since the 
following equality is \textit{not} always satisfied 
\bea
-(V(x) f_{\mu}(x)V(x+\mu)^\dag )^\dag
\equiv 
V(x+\mu) f_{\mu}(x)V(x)^\dag 
=V(x) f_{\mu}(x)V(x+\mu)^\dag ,
\nn\\
-(V(x) f_{\mu\nu}(x)V(x+\mu+\nu)^\dag)^\dag
\equiv V(x+\mu+\nu) f_{\mu\nu}(x)V(x)^\dag 
= V(x) f_{\mu\nu}(x)V(x+\mu+\nu)^\dag,\nn
\eea
due to the independence of gauge matrices $V(x)$ and $V(x+ \mu)$, 
$V(x+\mu+\nu)$.
Then such link and plaquette fields must be complexified
as $(C^a +i D^a) T^a$.
Therefore the new conjugate fields
$\eta^\dag, \phi^\dag, \phib^\dag, 
\psi^{\dag}_\mu, \chi^\dag, B^\dag_{12}$ are
\textit{independent} of 
$\eta$, $\phi$, $\phib$, $\psi_\mu$, $\chi$, $B_{12}$.\footnote{
It is not necessary to complexify the site fields
$\eta$, $\phi$, $\phib$. 
They can 
keep the anti-hermiticity under the gauge transformation
since they are in the adjoint representation.
Therefore $\eta^\dag = -\eta$,
$\phib^\dag = -\phib$ and $\phi^\dag =-\phi$ can be taken on the lattice.
Not only such a condition but also the condition $\eta^\dag = \eta$,
$\phib^\dag = \phib$ and $\phi^\dag =-\phi$ can be taken.}
Link gauge field $U_{\mu}$ are also complexified, which are not
unitary matrices, namely, 
\be
U_{\mu}(x) U^\dag_{\mu}(x) \ne 1.
\ee
They are defined as 
$U_\mu(x) = e^{A_\mu(x)}$ with
\textit{complexified} gauge fields $A_{\mu}(x)$ whose hermitian conjugate
$A^\dag_{\mu}(x)$ are not equal to $-A_{\mu}(x)$.

The continuum limit of the action~(\ref{caterall-act}) 
is \textit{different} from 
eq.~(\ref{gfermion}).
The 
degrees of freedom in eq.~(\ref{caterall-act}) are described with the general
complex matrices $\sum_a (C^a + i D^a)T^a$
while
the target theory~(\ref{gfermion})
is defined with anti-hermitian matrices
$\sum_a C^a T^a$.

%
%
%
\subsection{Correspondence between ${\cal N} = (4,4)$ CKKU lattice theory 
and Catterall's action}

If we neglect one ${\cal Q}$ multiplet, $\psi_{3,\nv} + \lambda_\nv$ 
and $\tilde{d}_\nv$, 
the CKKU's ${\cal N}$ =(4,4) lattice action~(\ref{orb-act-2})
and the ${\cal Q}$ transformations~(\ref{orb-BRST})
are same as the Catterall's action~(\ref{caterall-act})
and his $Q$ transformations~(\ref{cat-q}).
One can check the equivalence 
by identifying the fields as follows:
\be
\begin{array}{ll}
U_1(x) \LR \sqrt{2} z_{1,\nv}, & \psi_1(x) \LR \sqrt{2} \psi_{1,\nv},
\\
U_1^\dag(x) \LR \sqrt{2} \mybar z_{1,\nv}, 
& \psi_1^\dag (x)\LR - \sqrt{2} \xi_{2,\nv}
\\
U_2(x) \LR \sqrt{2} z_{2,\nv},  & \psi_2(x) \LR \sqrt{2} \psi_{2,\nv}
\\
U_2^\dag(x) \LR \sqrt{2} \mybar z_{2,\nv},  
& \psi_2^\dag(x) \LR \sqrt{2} \xi_{1,\nv},
\\
\bar{\phi}(x) \LR \sqrt{2} z_{3,\nv},  & \phi(x) \LR \sqrt{2} \mybar z_{3,\nv},
\\
\chi_{12}(x) \LR -i\sqrt{2} \chi_{\nv}, 
&\chi_{12}^\dag(x) \LR i\sqrt{2} \xi_{3,\nv}, 
\\
B_{12}(x) \LR -i\sqrt{2} \tilde{\mybar G}_{\nv},  
& B_{12}^\dag (x)\LR i\sqrt{2} \tilde{G}_{\nv}, 
\\
{\cal F}_{12}(x) \LR 2{\cal E}_\nv, 
& {\cal F}_{12}^\dag (x) \LR 2{\cal E}^{\dag}_\nv,
\\
\eta(x) \LR \sqrt{2} (\psi_{3,\nv} - \lambda_{\nv}), 
&
\end{array}
\label{correspo-catte}
\ee
where
\bea
{\cal E}_\nv = z_{1,\nv} z_{2,\nv+\xh}
-z_{2,\nv} z_{1,\nv+\yh} \nn \\
{\cal E}^\dag_\nv = \mybar z_{2,\nv+\xh} \mybar z_{1,\nv}
-\mybar z_{1,\nv+\yh} \mybar z_{2,\nv}.
\eea
Here, in eq.~(\ref{correspo-catte}),
the left hand sides of the symbol $\LR$ are the fields in 
Catterall's theory and the right hand sides are fields in CKKU's 
theory.\footnote{In eq.~(\ref{correspo-catte}),
we impose the condition \be
\phib^\dag = \phib, \,\, \eta^\dag = \eta, \,\, 
\phi^\dag = -\phi, 
\ee
on the site fields.}
Due to the complexification of the link and plaquette fields in 
``${\cal N} = (2,2)$'' Catterall's model, 
link fields $U_\mu(x)$,$U^\dag_\mu(x)$,~\textit{etc} of 
the Catterall's model can be identified
as the complex link fields $z_{i,\nv}$,$\mybar z_{i,\nv}$,~\textit{etc} 
of the ``${\cal N} = (4,4)$'' CKKU model
in the above correspondence.
Note that 
$\tilde{d}_\nv$ is a partner of
$\psi_{3,\nv} + \lambda_\nv$ 
under the 
${\cal Q}$ transformation
\be
{\cal Q}\tilde{d}_\nv =
i[\mybar z_{3,\nv},\psi_{3,\nv} + \lambda_\nv], \quad
{\cal Q}(\psi_{3,\nv} + \lambda_\nv) = -\sqrt{2} i\tilde{d}_\nv,
\ee
as in eq.~(\ref{orb-BRST}).
Other fields, except for $\mybar z_{3,\nv}$ whose ${\cal Q}$ transformation
is ${\cal Q} \mybar z_{3,\nv} = 0$,
do not appear in this transformation.
Therefore the absence of the set 
$\tilde{d}_\nv$ and $\psi_{3,\nv} + \lambda_\nv $
does not affect the ${\cal Q}$ transformation laws of other fields.
Moreover, since the set $\tilde{d}_\nv$ and $\psi_{3,\nv} + \lambda_\nv$ 
exists only in one term
\be 
{\cal Q}\sum_\nv \Tr \frac{1}{\sqrt{2}}(\psi_{3,\nv} + \lambda_\nv)
[i\tilde{d}_\nv -2
(\mybar z_{i,\nv -\ih}z_{i,\nv -\ih} -
z_{i,\nv}\mybar z_{i,\nv})]
\ee
among the terms of CKKU action~(\ref{orb-act-2}),
the action~(\ref{orb-act-2}) keep the ${\cal Q}$-exact form and the 
${\cal Q}$ symmetry 
under the truncation. 
As a side remark, correspondences 
among the symbol of lattice sites 
and the gauge coupling of both theories are
$x \LR \nv, \, -\beta \LR \frac{1}{2g^2}$.

%

%
%
\section{Relationship between Catterall model and Sugino's model}
As described in Section \ref{Sec:cat-act},
Catterall's ${\cal N} = (2,2)$ action has extra degrees of freedom 
which do not present in the target ${\cal N} =(2,2)$ theory.
Therefore 
it is necessary to 
truncate
the extra degrees of freedom 
to identify his model with an ${\cal N} = (2,2)$ lattice model 
which contains the correct number of degrees of freedom
of the target theory.
If this is performed in a naive way, breaking of the supersymmetry 
on the lattice would be resulted.
There exists a way of truncation which does not break the supersymmetry 
on the lattice.
Then, we find, after this truncation, that
the Catterall's theory 
\textit{
becomes the ${\cal N} =(2,2)$ supersymmetric lattice gauge theory 
being similar to
the Sugino model in ref.~\cite{Sugino:2003yb}.}%

\subsection{${\cal N} = (2,2)$ lattice model by Sugino}\label{Sec:Sugino-model}
To explain the derivation of the ${\cal N} = (2,2)$ Sugino type model
from the Catterall's ${\cal N} = (2,2)$ lattice action,
we briefly explain the Sugino's original ${\cal N} = (2,2)$ lattice 
model in \cite{Sugino:2003yb}.

His lattice action is 
\bea
S^{{\rm LAT}}_{{\cal N}=2} & = & Q\frac{1}{2g_0^2}\sum_x \, \Tr\left[ 
\frac14 \eta(x)\, [\phi(x), \,\bar{\phi}(x)] -i{\chi}(x){\Phi}(x) 
+{\chi}(x){H}(x)\right. \nn \\
 & & \hspace{2cm}\left. \frac{}{} 
+i\sum_{\mu=1}^d\psi_{\mu}'(x)\left(\bar{\phi}(x) - 
U_{\mu}(x)\bar{\phi}(x+\hat{\mu})U_{\mu}(x)^{\dagger}\right)\right], 
\label{lat_N=2_S}
\eea
where 
\be
\Phi(x)  =  -i\left[U_{12}(x)- U_{21}(x)\right],
\label{Phi_2d} 
\ee
and $U_{\mu\nu}(x)$ is plaquette variables
\be
U_{\mu\nu}(x) \equiv U_{\mu}(x) U_{\nu}(x+\hat{\mu}) 
U_{\mu}(x+\hat{\nu})^{\dagger} U_{\nu}(x)^{\dagger}. 
\quad (\mu,\nu = 1,2)
\ee
The target action of this model is the ${\cal N} = (2,2)$ super Yang-Mills 
action.
In this model,
the gauge fields are promoted to the compact unitary
variables
\be
U_\mu(x) = e^{ia A_\mu(x)}
\ee
on the link $(x,x+\mu)$. 
'$a$' stands for the lattice spacing, and $x \in \Z^2$.
All other fields sit on sites. 
Note that he uses the dimensionless variable in his paper.
The $Q$ transformations of this model are as follows
\bea
 & & QU_{\mu}(x) = i\psi_{\mu}'(x) U_{\mu}(x), \nn \\
 & & Q\psi_{\mu}'(x) = i\psi_{\mu}'(x)\psi_{\mu}'(x) 
    -i\left(\phi(x) - U_{\mu}(x)\phi(x+\hat{\mu})U_{\mu}(x)^{\dagger}\right),
  \nn \\
 & & Q\phi(x) = 0,     \nn \\
 & & Q\vec{\chi}(x) = \vec{H}(x), \quad 
           Q\vec{H}(x) = [\phi(x), \,\vec{\chi}(x)], \nn \\
 & & Q\bar{\phi}(x) = \eta(x), \quad  Q\eta(x) = [\phi(x), \,\bar{\phi}(x)]. 
\label{Q_lattice}
\eea
These $Q$-transformations satisfy following property
\be
Q^2 = (\text{infinitesimal gauge transformation with the parameter $\phi$
}).
\label{Q_nilpotent}
\ee
The action~(\ref{lat_N=2_S})
is invariant under the $Q$-transformation
since the action~(\ref{lat_N=2_S}) is
written by the $Q$-transformation of gauge invariant quantity.
After the $Q$-operation, the action~(\ref{lat_N=2_S})
takes the form 
\bea
S^{{\rm LAT}}_{{\cal N}=2} & = & \frac{1}{2g_0^2}\sum_x \, \Tr\left[
\frac14 [\phi(x), \,\bar{\phi}(x)]^2 + {H}(x){H}(x) 
-i{H}(x){\Phi}(x) \right. \nn \\
 & & \hspace{1.5cm}
+\sum_{\mu=1}^d\left(\phi(x)-U_{\mu}(x)\phi(x+\hat{\mu})U_{\mu}(x)^{\dagger}
\right)\left(\bar{\phi}(x)-U_{\mu}(x)\bar{\phi}(x+\hat{\mu})
U_{\mu}(x)^{\dagger}\right) \nn \\
 & & \hspace{1.5cm} -\frac14 \eta(x)[\phi(x), \,\eta(x)] 
- {\chi}(x) [\phi(x), \,{\chi}(x)] \nn \\
 & & \hspace{1.5cm}
-\sum_{\mu=1}^d\psi_{\mu}'(x)\psi_{\mu}'(x)\left(\bar{\phi}(x)  + 
U_{\mu}(x)\bar{\phi}(x+\hat{\mu})U_{\mu}(x)^{\dagger}\right) \nn \\
 & & \hspace{1.5cm}\left. \frac{}{}+ i{\chi}(x) Q{\Phi}(x) 
-i\sum_{\mu=1}^d\psi_{\mu}'(x)\left(\eta(x)-
U_{\mu}(x)\eta(x+\hat{\mu})U_{\mu}(x)^{\dagger}\right)\right]. \nn
\label{lat_N=2_S2}
\eea

%
%
%
%
\subsection{Derivation of the ${\cal N} =(2,2)$ Sugino type model by
a truncation of extra degrees of freedom in the Catterall's model} 
\label{Sec:cat-sugino}

In this subsection, we show that the Catterall's 
${\cal N} = (2,2)$ lattice model becomes 
${\cal N} = (2,2)$ lattice
model of the Sugino type 
if we truncate extra degrees of freedom 
in the Catterall's model 
by a way keeping supersymmetry on the lattice.

We start from the Catterall action
\begin{eqnarray}
S_L&=&-\beta Q{\rm Tr}
\sum_{x}\left(\frac{1}{4}\eta^\dagger(x)[\phi(x),\bar{\phi}(x)]-i
\chi^\dagger_{12}(x){\cal F}_{12}(x)
-i\chi_{12}(x){\cal F}_{12}(x)^\dagger\right.\nonumber\\
&+&
\biggl( \frac{1}{2}\chi^\dagger_{12}(x)B_{12}(x)+\frac{1}{2}\chi_{12}(x)B^\dagger_{12}(x) \nn \\
&&
+\frac{1}{2}\psi^\dagger_\mu(x)D^+_\mu\bar{\phi}(x)+\frac{1}{2}\psi_\mu(x)(D^+_\mu\bar{\phi}(x))^\dagger \biggr),
\label{caterall-act-2}
\end{eqnarray}
and the $Q$-transformation laws~(\ref{cat-q}).
\be
\begin{array}{ll}
QU_\mu=\psi_\mu &
QU^\dagger_\mu=\psi^\dagger_\mu,\nonumber\\
Q\psi_\mu=-D^+_\mu\phi, &
Q\psi^\dagger_\mu=-(D^+_\mu\phi)^\dagger,\nonumber\\
Q\chi_{12}=B_{12}, &
Q\chi^\dagger_{12}=B^\dagger_{12},\nonumber\\
QB_{12}=[\phi,\chi_{12}]^{(12)}, &
QB^\dagger_{12}=\left([\phi,\chi_{12}]^{(12)}\right)^\dagger,\nonumber\\
Q\phib=\eta, &
Q\phib^\dag=\eta^\dag,\nonumber\\
Q\eta=[\phi,\phib], &
Q\eta^\dag = ([\phi,\phib])^\dag, \nn \\
Q\phi=0. &
\end{array}
\label{cat-q2}
\ee
If we can possess the following property of 
$Q$-transformation~(\ref{nil-gauge});
\be
Q^2 = (\text{gauge transformation with parameter}\,\, \phi)
\ee 
even after truncation, we can keep supersymmetry under the  
truncation.

To perform such truncation,
we take the constraint $U_{\mu} (x) U_{\mu}^\dag (x) = 1$, namely 
$A_\mu^\dag(x) = -A_\mu(x)$
at first. Since $U_{\mu} (x) U_{\mu}^\dag (x) = 1$ is not dynamical quantity,
we obtain following conditions
\be
Q (U_{\mu}(x) U^\dag_{\mu}(x)) = 0, \quad
QU_{\mu}(x) = \psi_{\mu}(x), \quad Q U^\dag_\mu(x) = \psi^\dag_\mu(x).
\label{gauge-cons}
\ee
By this condition, 
$\psi^\dag_\mu(x)$ is described with $\psi_\mu(x)$ as
\be
\psi^\dag_\mu (x)= -U^\dag_\mu(x) \psi_{\mu}(x)U^\dag_\mu(x) ,
\label{psi-const}
\ee
and $\psi^\dag_\mu(x)$ is no longer independent of $\psi_\mu(x)$.
Then if we define a site fermion fields $\psi'_\mu(x)$ as
\be
\psi_{\mu}'(x) = \psi_\mu(x)U_\mu^\dag(x),\label{site}
\ee
two fermions $\psi_\mu(x)$ and $\psi_\mu^\dag(x)$ 
are described only by one fermion variable
$\psi_\mu'(x)$ as
\be
\psi_{\mu}(x) =  \psi_{\mu}'(x) U_{\mu}(x),
\quad
\psi^\dag_{\mu}(x) =  -U_{\mu}(x)^\dag\psi_{\mu}'(x).
\label{consist-psimu}
\ee
This $\psi_{\mu}'(x)$ becomes naturally \textit{anti-hermitian} since
\be
(\psi_{\mu}'(x))^\dag = (\psi_\mu(x)U_\mu^\dag(x))^\dag 
= U_\mu(x)\psi_\mu^\dag(x) = -\psi_\mu(x) U_\mu^\dag(x) = -\psi_{\mu}'(x).
\ee
This anti-hermitian property 
can be kept under the gauge symmetry since the site fields are adjoint 
representation. 
Then we take $\psi_{\mu}'(x)$ 
as a fundamental fermionic variable rather than $\psi_\mu(x)$. 
From this expression and the 
$Q$-transformation of $\psi_\mu$; $Q\psi_\mu(x) = \phi(x) 
U_\mu(x) - U_\mu(x)
\phi(x+ \mu)$, the $Q$ transformation law of $\psi_{\mu}'(x)$
is obtained naturally as
\bea
Q \psi_{\mu}'(x) &=& (Q\psi_{\mu}(x)U^\dag_\mu(x)) 
\nn \\
&=& (Q\psi_{\mu}(x))U^\dag_\mu(x) - \psi_{\mu}(x) (QU^\dag_\mu(x))
\nn \\
&=&  \psi_{\mu}'(x)\psi_{\mu}'(x) 
+ ( \phi(x) - U_\mu (x)\phi(x+ \mu) U^\dag_{\mu}(x)).
\label{transformation-psi}
\eea
These conditions $U_\mu(x)U^\dag_\mu(x)=1$ and the
eqs.~(\ref{gauge-cons}) -(\ref{transformation-psi}) 
give a way to truncate extra degrees
of freedom in 
gauge fields $U_\mu(x)$,$U_\mu^\dag(x)$ and their partners
$\psi_\mu(x)$, $\psi_\mu^\dag(x)$ without breaking of the supersymmetry on 
the lattice.

For $\phi(x)$, $\phib(x)$, $\eta(x)$, 
we impose 
$\eta^\dag(x) = -\eta(x)$,
$\phib^\dag(x) = -\phib(x)$ and
$\phi^\dag(x) = -\phi(x)$ to remove the extra degrees of freedom.
This condition can be kept under the gauge transformation since they are
in adjoint representation.
Since each $\phib$ and $\phib^\dag$
compose the $Q$-multiplets
with $\eta$ and $\eta^\dag$ respectively,
this condition does not break the supersymmetry on the lattice.

To truncate extra degrees of freedom in 
$\chi_{12}$, $\chi^{\dag}_{12}$, $B_{12}$
and~$B_{12}^{\dag}$
without breaking of the supersymmetry, 
we impose following constraints
\be
\begin{split}
\chi_{12}(x) &= \chi(x) U_1(x) U_2(x+1),  \\
\chi_{12}^\dag(x) &= - U_2^\dag(x+1) U_1^\dag(x) \chi(x),  \\
B_{12}(x) &= H(x) U_1(x) U_2(x+1) , 
 \\
B_{12}^\dag(x) &=  - U_2^\dag(x+1) U_1^\dag(x) H (x)
.
\end{split}
\label{substitute}
\ee
Here
$\chi(x)$ and~$H(x)$ are anti-hermitian site fields.
$\chi(x)$ and $H(x)$ 
are obtained by absorbing the link gauge fields $U_\mu(x)$ to 
$\chi_{12}(x)$ and $B_{12}(x)$
as $\chi(x) = \chi_{12}(x) U_2(x+1)^\dag U_1(x)^\dag$ and
$H(x) = B_{12}(x) U_2(x+1)^\dag U_1(x)^\dag$.
By the above condition,
$\chi_{12}^\dag(x)$ and~$B_{12}^\dag(x)$ are no longer independent of
$\chi_{12}(x)$ and~$B_{12}(x)$,
the degrees of freedom in these fields are reduced 
to two anti-hermite fields $\chi(x)$ and $H(x)$.
By performing the $Q$-transformation on right hand sides
and left hand sides of the above definitions
eq.~(\ref{substitute}), one can 
immediately
check that the $Q$ transformation
on $\chi(x)$, $H(x)$;
\bea
Q \chi(x) &=& H(x)
+ \chi(x) \psi_1'(x)
+ \chi(x) U_1(x) \psi_2'(x+1)U_1^\dag(x) 
\\
Q H(x) &=& \phi(x) \chi(x)
-\chi(x) U_1(x) U_2(x+1) \phi(x+1+2) U_2^\dag(x+1)U_1^\dag(x)
\nn \\
&& -H(x) \psi_1'(x) - H(x) U_1(x) \psi_2'(x+1)U_1^\dag(x)
\label{q-trns-cat}
\eea
is obtained consistently.
Note that $Q^2$ acts on these
$\chi(x)$ and $H(x)$ 
as the 
infinitesimal 
gauge transformation with the parameter $\phi$, 
namely
\be
Q^2 \chi(x) = [\phi(x), \chi(x)], \quad
Q^2 H(x) = [\phi(x), H(x)].
\ee

By these conditions,
\be
\frac{1}{2}\left( \chi_{12}^\dag(x) B_{12}(x) +
\chi_{12}(x) B_{12}^\dag(x) \right)
\ee 
becomes 
\be
-\chi (x)H(x).
\ee
One can check it by substituting eq.~(\ref{substitute})
to the Catterall action~(\ref{caterall-act-2}).
The term 
\be
-i\chi_{12}^\dag(x) \cF_{12}(x) -i
\chi_{12}(x) \cF_{12}^\dag(x)
\ee
becomes
\be
i\chi(x) \Phi(x)
\ee
where
\be
\Phi (x)= U_1(x)U_2(x+\hat{1})U_1^\dag(x+\hat{2})U_2^\dag(x)
-U_2(x)U_1(x+\hat{2})U_2^\dag(x+\hat{1})U_1^\dag(x).
\ee
Another term
\be
+\frac{1}{2}\psi^\dagger_\mu(x)D^+_\mu\bar{\phi}(x)+
\frac{1}{2}\psi_\mu(x)(D^+_\mu\bar{\phi}(x))^\dagger
\ee
becomes 
\be
-\psi_{\mu}'(x)(\bar{\phi}(x)-U_\mu
\bar{\phi} (x +\mu) U^\dag_\mu(x)).
\ee
It can also be checked
by substitution of eq.~(\ref{consist-psimu}) to the 
action~(\ref{caterall-act-2}).
Therefore,
Catterall's action~(\ref{caterall-act-2})
becomes 
\begin{eqnarray}
S_L&=&\beta Q{\rm Tr}\
\sum_{x}\Biggl(\frac{1}{4}\eta(x)[\phi(x),\bar{\phi}(x)]
+\chi(x)\biggl( H(x)-i\Phi(x) \biggr) \nn \\
&&
-\psi_{\mu}'(x)(\bar{\phi}(x)-U_\mu
\bar{\phi} (x +\mu) U^\dag_\mu(x)) \Biggr)
\label{sugino}
\end{eqnarray}
in the truncation. 
The $Q$-transformation laws are
\bea
Q U_\mu(x) &=& \psi_{\mu}' (x)U_\mu(x), \nn \\
 Q \psi_{\mu}'(x) &=& \psi_{\mu}'(x)\psi_{\mu}'(x)
+ ( \phi(x) - U_\mu (x)\phi(x+ \mu) U^\dag_{\mu}(x)), \nn \\ 
Q \chi(x) &=& H(x)
+ \chi(x) \psi_1'(x)
+ \chi(x) U_1(x) \psi_2'(x+1)U_1^\dag(x) 
\nn \\
Q H(x) &=& \phi(x) \chi(x)
-\chi(x) U_1(x) U_2(x+1) \phi(x+1+2) U_2^\dag(x+1)U_1^\dag(x)
\nn \\
&&
-H(x) \psi_1'(x) - H(x) U_1(x) \psi_2'(x+1)U_1^\dag(x)
\nn \\
Q \phib(x) &=& \eta (x), \nn \\
Q \eta(x) &=& [\phi(x),\phib (x)],
\nn \\
Q \phi(x) &=& 0 .
\label{cat-rest-trans}
\eea
The $Q$-transformation laws~(\ref{cat-q2})
become eq.~(\ref{cat-rest-trans}) by the truncation.
After the $Q$-operation, 
this action (\ref{sugino}),
becomes
\bea
S_L &=& \beta \sum_x \Tr \Biggl(
\frac{1}{4} [\phi(x), \phib(x)]^2
+H(x)(H(x) -i \Phi(x) )
 \nn \\
&& -\psi_{\mu}'(x)\psi_{\mu}'(x)(\bar{\phi}(x)+U_\mu
\bar{\phi} (x +\mu) U^\dag_\mu(x)) 
\nn \\
&&-(\phi(x)-U_\mu
\phi (x +\mu) U^\dag_\mu(x))
(\bar{\phi}(x)-U_\mu
\bar{\phi} (x +\mu) U^\dag_\mu(x))
\nn \\ 
&& -\frac{1}{4}\eta(x)[\phi(x),\eta(x)]
\nn \\
&&
-\chi(x)\left( \phi(x)\chi(x)
-\chi(x) U_1(x)U_2(x+1) \phi(x+1+2) U_2^\dag (x+1) U_1^\dag(x)
\right)
\nn \\
&& +i \chi(x) U_1(x)U_2(x+1)Q(U_2^\dag(x+1) U_1^\dag(x)
-U_1^\dag(x+2)U_2(x))
\nn \\
&& -i \chi(x) Q (U_1(x)U_2(x+1)-U_2(x)U_1(x+2))
U_2^\dag(x+1) U_1^\dag(x)
\nn \\
&& +\psi_{\mu}'(x)(\eta(x)-U_\mu
\eta (x +\mu) U^\dag_\mu(x)) 
\Biggr).
\label{sugino-after-Q2}
\eea
This action 
eqs.~(\ref{sugino}),(\ref{sugino-after-Q2})
has a correct continuum limit eq.~(\ref{gfermion})
while the original Catterall action~(\ref{caterall-act-2}) does not have.

Note that the action~(\ref{sugino}),(\ref{sugino-after-Q2})
is almost same as 
Sugino's action~(\ref{lat_N=2_S}) and (\ref{lat_N=2_S2}).
Only the fermionic terms
\be
-\chi(x)\left( \phi(x)\chi(x)
-\chi(x) U_1(x)U_2(x+1) \phi(x+1+2) U_2^\dag (x+1) U_1^\dag(x)
\right)
\ee
and 
\bea
&+i \chi(x) U_1(x)U_2(x+1)Q(U_2^\dag(x+1) U_1^\dag(x)
-U_1^\dag(x+2)U_2(x)) \nn \\
&-i \chi(x) Q (U_1(x)U_2(x+1)-U_2(x)U_1(x+2))
U_2^\dag(x+1) U_1^\dag(x)
\eea
are different from their corresponding terms
$-\chi(x)[\phi(x), \chi(x)]$ and
$i\chi(x)Q \Phi(x)$ in Sugino's original 
model~(\ref{lat_N=2_S}),(\ref{lat_N=2_S2}).
After the integration over the auxiliary field $H(x)$,
the gauge kinetic term 
\be
-\beta \sum_x\Tr \frac{1}{4}\Phi^2 (x)= -\beta \sum_x\Tr \frac{1}{4}
(U_{12}(x) -U_{21}(x))^2
\ee
is obtained.
This is same as the gauge kinetic term in Sugino's original model.
Therefore, also the action~(\ref{sugino}),(\ref{sugino-after-Q2})
has the vacuum degenerate problem which
the original Sugino model encountered in ref.~\cite{Sugino:2003yb}.
\footnote{Sugino has proposed several treatments to solve this problem in
refs.~\cite{Sugino:2003yb,Sugino:2004qd}.}  

Also the $Q$-transformation laws 
after the truncation~(\ref{cat-rest-trans})
are almost same as
the $Q$-transformation laws
of the Sugino's model~(\ref{Q_lattice}).
Only the transformation laws of 
auxiliary field $H(x)$ and its partner
$\chi(x)$ in (\ref{cat-rest-trans})
are different from the $Q \chi(x) = H(x)$ and
$Q H(x) = [\phi(x), \chi(x)]$ of Sugino's original model.

As a result, \textit{Catterall's model becomes
the theory which is almost same as
the Sugino's theory 
by the truncation of extra degrees of freedom 
which does not break supersymmetry}.

\section{Relationship between Sugino's model and CKKU model}
Due to the two relationships
described in section 2 and 3,
it is obvious
that the model of the Sugino type can be derived by the truncation 
of degrees of freedom in the CKKU model.
Due to
the relationship 
between CKKU model and Catterall's model,
the method to derive the Sugino type model from the Catterall's theory is 
applicable to derive the model of the Sugino type from CKKU model.
Since the explanation of the derivation is mere repetition of
the description in the subsection~\ref{Sec:cat-sugino},
we put off the explanation of the derivation 
in the Appendix A.

In this section,
we explain that the derivation
discards the 
fluctuations along the flat-direction around the 
vacuum expectation value $\frac{1}{\sqrt{2}a}$ of scalar potential 
existing in the CKKU model.\footnote{In this section, 
we take into account the the flat-directions
of moduli space
while we neglect such effects in section 2; 
we showed the relationship between the CKKU model
and the Catterall model without the consideration of such effects.}
 
To explain it, we explain the deconstruction and 
the fluctuation in the CKKU model
at first.
Then we explain that the derivation truncates such fluctuations.

\subsection{The deconstruction and 
the fluctuations along the moduli space in CKKU model}

To realize the kinetic term in CKKU model,
performing the ``deconstruction'' is required.(see also section 3.3 in 
ref.~\cite{Cohen:2003qw}.)
The deconstruction is the field redefinition of the bosonic link fields
$z_{i,\nv}$ expanding around the vacuum expectation value 
\be
\langle z_{i,\nv} \rangle = \frac{1}{\sqrt{2}a} \mathbf{1}_M,
\ee
where the $\mathbf{1}_M$ is $M \times M$ unit matrix and
the $a$ is interpreted as lattice spacing.

To perform the expansions,
there are two ways of representations;
Cartesian decomposition and the polar decomposition.
These two decomposition give the same continuum limit
as Unsal proved in ref.~\cite{Unsal:2005yh}.
CKKU adopts the Cartesian decomposition, eq.(3.16) in 
ref.~\cite{Cohen:2003qw}, which represents 
the complex link variables
by the sum of hermitian matrices and the antihermitian matrices.
But, to perform the derivation of the Sugino type model,
we have to adopt the 
\textit{polar decomposition}~\cite{Unsal:2005yh,Onogi:2005cz,Ohta:2006qz}.

In the polar decomposition,
the bosonic link fields
$z_i$,$\mybar z_i \, (i = 1,2)$ 
are uniquely represented as a product of hermitian
matrices $(\frac{1}{a} + s_{i,\nv})$ ($i =1,2$),
which represent a radial direction and so have positive eigenvalues only, 
and unitary matrices $U_{i,\nv}$ 
\be
\begin{array}{ll}
z_{i,\nv}=\frac{1}{\sqrt{2}}
\left(\frac{1}{a}{\bf 1}_M+s_{i,\nv}\right)U_{i,\nv},&
\mybar z_{i,\nv}=\frac{1}{\sqrt{2}}
U_{i,\nv}^\dag\left(\frac{1}{a}{\bf 1}_M+s_{i,\nv}\right),\\
\end{array}
\label{unitaryxy2-2}
\ee
where lattice spacing $\frac{1}{\sqrt{2}a}$ and the scalar
fields $s_{i,\nv}$ sit on sites
and $U_{i,\nv}$ are link fields
written by the gauge fields $v_{i,\nv}$ as
$U_{i,\nv} = e^{iav_{i,\nv}}$. 
Comparing with the Cartesian decomposition 
in ref.~\cite{Cohen:2003qw},
this representation of decomposition has advantage of the manifest gauge
symmetry.
This representation is required to
keep the gauge symmetry under the truncation.

Note that, in the CKKU model,
the lattice spacing is dynamical quantity 
characterized as the vacuum expectation value 
$\frac{1}{\sqrt{2}a}$
of scalar potential. 
The scalar fields 
$s_i$ are fluctuations around the $\frac{1}{\sqrt{2}a}$.

The CKKU action has noncompact moduli space consisting 
of all constant scalar fields satisfying $[s_1,s_2]=0$.
The integral of these modes are formally divergent,
the expansion~(\ref{unitaryxy2-2}) is then poorly defined.
(Even if we take the Cartesian decomposition taken in 
ref.~\cite{Cohen:2003qw}, such instability of the 
vacuum occurs.)
To suppress the large fluctuation along the flat directions,
the original CKKU model introduced the moduli fixing mass term
\be
\sum_{\nv} \Tr 
\left[\left( z_{i,\nv} \mybar z_{i,\nv} - \frac{1}{2a^2} \right)^2\right]
= \sum_{\nv}\Tr \frac{1}{4}
\left[\left( (s_{i,\nv} + \frac{1}{a})^2 - 
\frac{1}{a^2} \right)^2\right].
\label{modu-fix}
\ee

\subsection{Truncation of the flat-direction by the derivation
of the Sugino type model}

When we derived the model of Sugino type from the Catterall model,
we imposed the condition
that the link variables
become unitary, namely,
\be
U_\mu(x)U^\dag_{\mu}(x) = 1.
\ee 
Therefore, from the correspondence between 
the fields of CKKU model and the ones of Catterall 
model~(\ref{correspo-catte}),
complex link fields 
$z_{i,\nv}$ in the CKKU model become ``unitary'' link variables
to derive the model of the Sugino type.
This means that dynamical degrees of freedom which correspond to 
radial directions of the links $z_{i}$, $\mybar z_i$
are discarded in the derivation,
namely,
\be
z_{i,\nv}=\frac{1}{\sqrt{2}a}U_{i,\nv}, \quad 
\mybar z_{i,\nv} = \frac{1}{\sqrt{2}a}U^\dag_{i,\nv},
\ee
where the vacuum expectation value $\frac{1}{\sqrt{2}a}$ 
cannot be removed
since the link fields $z_i, \mybar z_i$ have 
mass dimension 1.

Note that the derivation of Sugino type model from CKKU model
discards the fluctuations $s_i$ around 
the vacuum expectation value $\frac{1}{\sqrt{2}a}$.
Then, also the large fluctuations of $s_i$ along the flat-directions
which cause the serious instability of the vacuum 
are removed under the derivation.
Therefore we do not have to introduce the moduli fixing mass term
in the derived Sugino type model.
Moreover we can derive the model of the Sugino type from the CKKU model
even if we introduce the moduli fixing mass 
term~(\ref{modu-fix}),
\be
\sum_{\nv} \Tr 
\left[\left( z_{i,\nv} \mybar z_{i,\nv} - \frac{1}{2a^2} \right)^2\right]
= \sum_{\nv}\Tr \frac{1}{4}
\left[\left( (s_{i,\nv} + \frac{1}{a})^2 - 
\frac{1}{a^2} \right)^2\right]
\ee
in the CKKU model.
This is because the mass term naturally vanishes under the truncation
$s_{i,\nv} = 0$.

\section{Corresponding truncation in the continuum theory}
\label{Sec:cont-trunc}
The derivation of the Sugino type model
from the CKKU model (or Catterall model)
can be interpreted as 
the lattice analogue of the derivation of 
the continuum ${\cal N} = (2,2)$ theory
from the continuum ${\cal N}= (4,4)$ theory 
by the truncation of several $Q$-multiplets.

We first consider the continuum ${\cal N} = (4,4)$ supersymmetric gauge theory
action 
\be
S=\frac{1}{g_2^2}\displaystyle \int d^2x 
Q\Xi
\label{TFT-cont2-1}
\ee
where
\be
\begin{split}
\Xi
&= \Tr\Bigg[\frac{1}{4}\eta[\phi,\phib]
+\chi^\R
(H^\R-i{\cal E}^\R)
+\chi_1
(H_1-i{\cal E}_1)
+\chi_2
(H_2-i{\cal E}_{2})
\\
&\qquad+\frac{1}{2}
\biggl\{
\psi_{\mu}
D_{\mu} \phib 
+\psi_{s_i} [ s_i, \phib] 
\biggr\}\Bigg],
\end{split}
\label{orb-act-3}
\nn
\ee
and
\bea
{\cal E}^\R &=& -2(D_1 s_1 + D_2 s_2),\nn\\
{\cal E}_1 &=& 2(D_1 s_2 - D_2 s_1),\nn \\
{\cal E}_2 &=& 2(i[s_1,s_2] + F_{12}),\nn
\\
F_{12} &=& -i[D_1,D_2].\nn
\eea
Here the indices $\mu$, $i$ run from 1 to 2,
and the repeated indices are summed.
$s_i$, $\phi$, $\phib$ 
are bosonic scalar fields and 
$H^\R$, $H_i$ are auxiliary fields and $v_\mu$ are gauge fields.
The others 
$\psi_{s_i}$, $\psi_\mu$, $\chi^\R$, $\chi_i$
are fermionic fields.
The all fields are in 
adjoint representation
of the gauge group.
$D_\mu$ is the covariant derivative.
$Q$-transformation laws are
\be
\begin{array}{ll}
Qs_i = (\psi_{s_i}),
&
Q\psi_{s_i} = [\phib, s_i],
\\
Q\phib = \eta,
&
Q\eta = [\phi, \phib],
\\
Qv_\mu = \psi_\mu,
&
Q \psi_\mu = iD_\mu \phi ,
\\
Q\chi^{\R} = H^\R,
&
QH^\R = [\phi, \chi^\R],
\\
Q\chi_{i} = H_i
&
QH_i = [\phi, \chi_i]\, (i = 1,2),
\\
Q\phi = 0.
&
\end{array}
\label{q-definition}
\ee
In each $Q$-transformation law of each $Q$-multiplet,
only the components of the multiplet and 
$\phi$, whose transformation is $Q\phi = 0$,
appear. 
Also note that the twice operation of $Q$
generates the infinitesimal gauge transformation 
with parameter
$\phi$.
The action is invariant under the $Q$-operation since 
it is written as
$Q$-transformation of the gauge invariant quantity.

To derive the ${\cal N} = (2,2)$ supersymmetric theory from the 
${\cal N} = (4,4)$ theory, we discard following
$Q$-multiplets;
\begin{enumerate}
\item $\chi^\R$ and $H^\R$, which 
are contained only in the term $\chi^R(H^\R -i {\cal E}^\R)$ among the terms in
$\Xi$ of eq.~(\ref{TFT-cont2-1})
\item $\chi_1$ and $H_1$ contained only in the term 
$\chi_1(H_1 -i {\cal E}_1)$ 
\item $s_i$ and $\psi_{s_i}$ contained only in
$\frac{1}{2} \psi_{s_i}[s_i,\phi]$ and 
$2\chi_2[s_1,s_2]$.
\end{enumerate}
If we substitute the condition 
\be
s_i = \psi_{s_i}= \chi^\R = H^\R = H_1 =\chi_1 =0,
\label{truncate-23}
\ee
the action~(\ref{TFT-cont2-1})
reduces to 
\be
S_{(2,2)}
=\frac{1}{g_2^2}\displaystyle \int d^2x 
Q \Xi'
\label{TFT-cont2-2}
\ee
\be
\Xi' = \Tr\Bigg[\frac{1}{4}\eta[\phi,\phib]
+\chi
(H-i{\cal E})
+\frac{1}{2}
\biggl\{
\psi_{\mu}
D_{\mu} \phib 
\biggr\}\Bigg]
\nn
\ee
where
\bea
{\cal E} &=& 2(F_{12}),\nn
\\
F_{12} &=& -i[D_1,D_2].\nn
\eea
$\Xi$ in eq.~(\ref{TFT-cont2-1})
reduces to the above gauge invariant quantity $\Xi'$ under the
truncation.
Note that the absence of the three $Q$-multiplets 
$s_i$ and $\psi_{s_i}$, $etc$ 
does not change the 
$Q$-transformations of other fields.
This is because only the remaining fields $\eta$, $\phib$, $etc$,
which survive under the truncation,
appear in the 
$Q$-transformation laws of the remaining fields.
Moreover the condition~(\ref{truncate-23}) is kept under the
$Q$-transformation.
Also note
this action~(\ref{TFT-cont2-2}) 
is also written as the $Q$-operation on the gauge invariant quantity
$\Xi'$.
Therefore the action~(\ref{TFT-cont2-2})
keeps the $Q$-symmetry.
This action~(\ref{TFT-cont2-2}) is equivalent to 
the continuum ${\cal N} = (2,2)$ supersymmetric gauge theory.
Finally, we obtain the ${\cal N} = (2,2)$ theory by the truncation of 
degrees of freedom in the ${\cal N} = (4,4)$ theory.

The derivation of the Sugino type model
from the CKKU model (or Catterall model) 
is the lattice analogue of the derivation of this 
continuum ${\cal N} = (2,2)$
theory~(\ref{TFT-cont2-2})
from ${\cal N} = (4,4)$ theory.

\section{Conclusion}
In this paper,
we clarified the relationship between several,
seemingly quite different,
supersymmetric lattice models
preserving supersymmetry on the lattice.
First we showed that Catterall's model can be embedded in CKKU's model
as a sub-sector. 
Also we clarified that a model of the 
Sugino type naturally appears 
when we truncate the 
degrees of freedom in Catterall's model
in a way which does not break the supersymmetry on the lattice.
We also show that 
the ${\cal N} = (4,4) $ CKKU model can give the Sugino type model
if we truncate the fluctuations 
around the vacuum expectation value $\frac{1}{\sqrt{2}a}$
and other degrees of freedom.

These relationships would indicate an underlying essential 
structure which any lattice formulations preserving partial 
supersymmetry possess.
Further understanding of this structure would be very useful
to develop lattice formulations of 
supersymmetric gauge theory.

Since the Catterall's lattice model and the model of Sugino type can be 
built also from the CKKU lattice model which is constructed from 
super matrix model,
we would be able to utilize the super matrix model analysis for 
these lattice formulations.
There is a possibility 
that also Catterall's model and the Sugino type model could be 
described using the matrix model analysis.
\\
\\
{\bf \large Acknowledgment}
\\
The author would like to thank to P.~Damgaard, Y.~Kikukawa,
S.~Matsuura, K.~Ohta,
T.~Onogi,
H.~Suzuki and  M.~Unsal
for valuable discussions and comments.
In particular, he would like to express his
gratitude to S.~Matsuura for 
crucial comments
on the previous version of the paper.

\appendix
\section{Derivation of the ${\cal N} =(2,2)$ Sugino type model from CKKU model}
We will explicitly show the derivation of  
the Sugino type model from
the ${\cal N} =(4,4)$ CKKU lattice model.
Here we utilize the technology 
to derive the ${\cal N} = (2,2)$ lattice theory from ${\cal N} =(4,4)$ 
lattice theory proposed in ref.~\cite{Ohta:2006qz}.

At first, we 
truncate some scalars and auxiliary fields
by imposing $\tilde{d}_\nv =  s_i= \psi_{3,\nv} + \lambda_\nv = 0$.
After this truncation, 
the expansion eq.~(\ref{unitaryxy2-2})
of the bosonic link fields 
$z_{i,\nv}$, $\mybar z_{i,\nv}$
become
\be
\begin{array}{ll}
z_{i,\nv} =
\frac{1}{\sqrt{2}a} U_{i,\nv}, &
\mybar z_{i,\nv} =
\frac{1}{\sqrt{2}a} U_{i,\nv}^\dag.
\label{truncation}
\end{array}
\ee
Note that $\mybar z_{i,\nv}$ are no longer independent of $z_{i,\nv}$ 
due to the absence of the scalar fields $s_{i,\nv}$.
Since only the scalar fields $s_i$ can give the dynamical fluctuations 
and the radiative corrections
of lattice spacing, 
the lattice spacing $\frac{1}{\sqrt{2}a}$ is no longer dynamical
quantity.
The product of these two link fields $z_{i,\nv}$ and the 
$\mybar z_{i,\nv}$
combine the \textit{non-dynamical}
lattice spacing as $z_{i,\nv}\mybar z_{i,\nv} = \frac{1}{2a^2}$.
Therefore, we can take a condition that the 
${\cal Q}$-transformation of the 
product $z_{i,\nv}\mybar z_{i,\nv} = \frac{1}{2a^2}$
vanishes.
Thus, from eq.~(\ref{truncation}), 
we immediately obtain 
\be
\begin{split}
&{\cal Q} z_{i,\nv} = \frac{1}{\sqrt{2}a} {\cal Q} U_{i,\nv} = \psi_{i,\nv},\qquad
{\cal Q} \mybar z_{i,\nv} =  \frac{1}{\sqrt{2}a} {\cal Q} U_{i,\nv}^\dag = - \epsilon^{ij}
\xi_{j,\nv}, \\
&{\cal Q} (z_{i,\nv} \mybar z_{i,\nv}) = {\cal Q} \frac{1}{2a^2} = 0.
\label{constraints for XX}
\end{split}
\ee
From eqs.~(\ref{truncation}),(\ref{constraints for XX}),
we obtain the constraints between 
fermions $\psi_i$ and $\epsilon_{ij}\xi_j$
\be
\begin{split}
&-\epsilon_{ij}\xi_{j,\nv} = - U^\dag_{x,\nv} \psi_{i,\nv} U^\dag_{x, \nv}. \\
\end{split} \label{lambda-condition}
\ee
By this definition, half of degrees of freedom in 
complex fermion fields $\psi_{i,\nv}$ and
$\xi_{i,\nv}$ are discarded. $\xi_{i,\nv}$ are no longer independent of 
$\psi_{i,\nv}$.
Due to the relationships~(\ref{lambda-condition}),
we can represent the above link fermions $\psi_{i,\nv}$,
$\xi_{i,\nv}$
by absorbing the
link variables as
\bea
&\psi_{i,\nv} = i \psi^{i}_\nv U_{i,\nv}, \\
&-\epsilon_{ij}\xi_{j,\nv} = -i U_{i,\nv}^\dag \psi^{i}_\nv ,
\eea
where $\psi^i_\nv$ are site fermions in the adjoint representation.
The ${\cal Q}$-transformations of the site fermions $\psi^i_{\nv}$ are
naturally obtained as
\be
{\cal Q} \psi^i_\nv
=
\sqrt{2}a \,\psi^i_\nv\psi^i_\nv -i \frac{1}{a}\left(
\mybar z_{3,\nv} - U_{i,\nv} \mybar z_{3,\nv+\iv} U^\dag_{i,\nv}
\right).
\ee

We can 
also discard the half of degrees of freedom 
in the fermion fields
$\chi_\nv$ and $\xi_{3,\nv}$
by imposing the condition 
\bea
&\chi_{\nv} = - \chi'_{\nv}U_{1,\nv} U_{2,\nv+ \xh}\label{1},\\
&\xi_{3,\nv} = - U^\dag_{2,\nv+ \xh} U^\dag_{1,\nv} \chi'_{\nv}\label{2},
\eea
where $\chi'_\nv$ is a site fermion.
By this condition, $\chi_{\nv}$ is no longer independent of $\xi_{3,\nv}$.
The
truncation of the degrees of freedom in 
the bosonic auxiliary fields 
$\tilde{\mybar G}_\nv, \tilde{G}_\nv $
are also performed by 
absorbing
the link fields as,
\bea
\tilde{G}_\nv 
= {\cal Q} \xi_{3,\nv} 
& \equiv &
-U^\dag_{2,\nv+ \xh} U^\dag_{1,\nv} H_\nv
\nn \\
&=& -(\sqrt{2} a \xi_{1,\nv+ \xh} U^\dag_{1,\nv} \chi'_\nv
-\sqrt{2} a  U^\dag_{2,\nv+ \xh} \xi_{2,\nv} \chi'_\nv
+ U^\dag_{2,\nv+ \xh} U^\dag_{1,\nv} {\cal Q} \chi'_\nv),
\nn \\
\tilde{\mybar G}_\nv 
= {\cal Q} \chi_{\nv}
& \equiv &
-H_\nv U_{1,\nv} U_{2,\nv+ \xh}
\nn \\
&=& -({\cal Q} \chi'_\nv U_{1,\nv} U_{2,\nv+ \xh}
- \sqrt{2} a \chi'_\nv \psi_{1,\nv} U_{2,\nv+ \xh}
- \sqrt{2} a \chi'_\nv U_{1,\nv} \psi_{2,\nv+ \xh}
).
\eea
where $H_\nv$ is a bosonic site field. 
The ${\cal Q}$-transformation laws of $\chi'_\nv$ 
and $H_\nv$ are
\bea
{\cal Q} \chi'_\nv &=& H_\nv
+i \sqrt{2} a \psi^1_\nv \chi'_\nv
+i \sqrt{2} a U_{1,\nv} \psi^2_{\nv+\xh} U_{1,\nv}^\dag \chi'_\nv ,
\nn \\
{\cal Q} H_\nv &=& \sqrt{2}
\Biggl(
-( \chi'_\nv \mybar z_{3,\nv}
- U_{1,\nv}U_{2,\nv+\xh} \mybar z_{3,\nv+\xh+\yh}
U_{2,\nv+\xh}^\dag U_{1,\nv}^\dag \chi'_\nv)
\nn \\
&& + i a U_{1,\nv} \psi^2_{\nv+\xh} U_{1,\nv}^\dag
H_\nv
+ i a \psi^1_\nv H_\nv
\Biggr)
.
\label{q-trans}
\eea
The above conditions 
eqs.~(\ref{constraints for XX})-(\ref{q-trans})
in ${\cal N} = (4,4)$
CKKU lattice theory are almost same as the truncation conditions 
eqs.~(\ref{gauge-cons})-(\ref{q-trns-cat}) which derive the model 
of the Sugino type
from
Catterall's model in the subsection \ref{Sec:cat-sugino}.
Then, the property 
\be
{\cal Q}^2 = (\text{infinitesimal gauge transformation with parameter 
$\mybar z_3$})
\ee
is kept even after the truncations.
Therefore, the ${\cal N} = (4,4)$ CKKU lattice action 
can be truncated to 
${\cal N} = (2,2)$ lattice action 
with a preserved supercharge ${\cal Q}$.
The ${\cal N} = (2,2)$ lattice action is described as
follows,
\be
S = \frac{1}{2g^2} {\cal Q} \, \Xi'
\label{trun-act}
\ee
where
\be
\begin{split}
\Xi'
&= \sum_{\nv}
\Tr\Bigl[\frac{1}{\sqrt{2}}(\psi_{3,\nv} - \lambda_\nv)
[\mybar z_{3, {\bf n}},z_{3, {\bf n}}]
\\
&\qquad+2
\chi'_\nv H_\nv
-i \frac{\sqrt{2}}{a^2}\chi'_\nv ( \Phi_\nv)
+\frac{2i}{a}
\psi^i_\nv(z_{3,\nv}
-U_{i,\nv}z_{3,\nv+\ih} U^\dag_{i,\nv}) \Bigr],
\\
&
\Phi_\nv = -i \left( U_{1,\nv}U_{2,\nv +\xh}U^\dag_{1, \nv+ \yh}U^\dag_{2,\nv}
- U_{2,\nv}U_{1,\nv +\yh}U^\dag_{2,\nv+ \xh}U^\dag_{1,\nv} \right).
\end{split}
\label{tru-act-2}
\ee
The ${\cal Q}$ transformations of the fields in eq.~(\ref{tru-act-2})
are summarized as
\bea
 & & {\cal Q}U_{i,\nv} = i\psi^{i}_\nv U_{i,\nv}, \nn \\
 & & {\cal Q}\psi^{i}_{\nv} = \sqrt{2}a\psi^{i}_{\nv}\psi^{i}_{\nv} 
    -i\frac{1}{a}\left(\mybar z_{3,\nv} 
- U_{i,\nv}\mybar z_{3,\nv+ \ih}U_{i,\nv}^{\dagger}\right),
  \nn \\
 & & {\cal Q}\mybar z_{3,\nv} = 0,     \nn \\
 & & {\cal Q}\chi'_\nv = H_\nv 
+i \sqrt{2} a \psi^1_\nv \chi'_\nv
+i \sqrt{2} a U_{1,\nv} \psi^2_{\nv+\xh} U_{1,\nv}^\dag \chi'_\nv ,
\nn \\
& & {\cal Q} H_\nv = 
\sqrt{2}
\Biggl(
-( \chi'_\nv \mybar z_{3,\nv}
- U_{1,\nv}U_{2,\nv+\xh} \mybar z_{3,\nv+\xh+\yh}
U_{2,\nv+\xh}^\dag U_{1,\nv}^\dag \chi'_\nv)
\nn \\
& & \qquad \qquad
+ i a U_{1,\nv} \psi^2_{\nv+\xh} U_{1,\nv}^\dag
H_\nv
+ i a \psi^1_\nv H_\nv
\Biggr),
\nn \\
 & & {\cal Q}z_{3,\nv} = \psi_{3,\nv}-\lambda_\nv, \quad  
{\cal Q}(\psi_{3,\nv}-\lambda_\nv) 
= \sqrt{2}[\mybar z_{3,\nv}, \,z_{3,\nv}]. 
\label{Q_lattice02}
\eea
After the ${\cal Q}$-operation, 
the action becomes
\bea
S & = & \frac{1}{2g^2}\sum_\nv \, \Tr\left[
 [ \mybar z_{3,\nv}, \, z_{3,\nv}]^2 
+ 2  H_\nv H_\nv 
-i\frac{\sqrt{2}}{a^2} H_\nv \Phi_\nv \right. \nn \\
 & & \hspace{1.5cm}
+\sum_{i=1}^2 \frac{2}{a^2} \left( \mybar z_{3,\nv}-U_{i,\nv}
\mybar z_{3,\nv+\ih}
U_{i,\nv}^{\dagger}
\right)
\left( z_{3,\nv}
- U_{i,\nv}
z_{3,\nv+\ih}
U_{i,\nv}^{\dagger}
\right) \nn \\
 & & \hspace{1.5cm} -\frac{1}{\sqrt{2}} 
(\psi_{3,\nv}-\lambda_\nv)
[ \mybar z_{3,\nv}, \,(\psi_{3,\nv}-\lambda_\nv)] 
\nn \\
%
%
&& \hspace{1.5cm}
-2\sqrt{2}  \chi'_\nv  (\mybar z_{3,\nv}  \chi'_\nv 
-\chi'_\nv U_{1,\nv}^\dag U_{2,\nv+\xh} \mybar z_{3,\nv+\xh+\yh}
U_{2,\nv+\xh}^\dag U_{1,\nv}^\dag)
\nn \\
%
%
 & & \hspace{1.5cm}
-\sum_{\mu=1}^2
2\sqrt{2}  \psi^i_{\nv} \psi^i_{\nv} \left( z_{3,\nv } + 
U_{i,\nv} z_{3,\nv+\xh} U_{i,\nv}^{\dagger}\right) \nn \\
 & & \hspace{1.5cm}
%
%
+  \frac{\sqrt{2}}{a^{2}} \chi'_\nv U_{1,\nv}U_{2,\nv+\xh}
{\cal Q}(U_{1,\nv+\yh}^\dag U_{2,\nv}^\dag
- U_{2,\nv+\xh}^\dag U_{1,\nv}^\dag)
\nn \\
 & & \hspace{1.5cm}
+\frac{\sqrt{2}}{a^{2}} \chi'_\nv {\cal Q} 
(U_{1,\nv}U_{2,\nv+\xh} -U_{2,\nv}U_{1,\nv+\yh})
U_{2,\nv+ \xh}^\dag U_{1,\nv}^\dag
\nn \\
%
%
 & & \hspace{1.5cm}\left. \frac{}{}
-i\sum_{i=1}^2 \frac{2}{a} \psi^i_\nv
\left((\psi_{3,\nv}-\lambda_\nv)
- U_{i,\nv} (\psi_{3,\nv+\ih}-\lambda_{\nv+\ih})
U_{i,\nv}^{\dagger}\right)\right]. 
\label{CKKU=Q-act}
\eea

One can confirm that 
the ${\cal Q}$-transformation~(\ref{Q_lattice02}) and 
the action~(\ref{tru-act-2})(\ref{CKKU=Q-act}) are almost same as 
$Q$-transformation in the Sugino's model~(\ref{Q_lattice}) and 
his action~(\ref{lat_N=2_S}),(\ref{lat_N=2_S2}),
by following identifications,
\be
\begin{array}{ll}
\eta(x) \LR \sqrt{2}a^{3/2}(\psi_{3,\nv} - \lambda_\nv),
& \phib(x) \LR \sqrt{2}a \,z_{3,\nv},
\nn \\
\chi(x) \LR \sqrt{2} a^{3/2} \chi'_\nv,
&
H(x) \LR \sqrt{2} a^{2} H_\nv,
\nn \\
U_{\mu} (x) \LR U_{i,\nv},
&
\psi_{\mu}' (x) \LR \sqrt{2} a^{3/2} \psi^{i}_\nv,
\nn \\
\phi(x) \LR \sqrt{2}a \,\mybar z_{3,\nv},
& \nn \\
\mu \LR i,
&
Q \LR a^{1/2}{\cal Q},
\nn \\
\frac{1}{2g_0^2} \LR \frac{1}{2g^2}a^{-4}.
\end{array}
\ee
Only the several fermionic terms in eq.~(\ref{CKKU=Q-act})
\bea
%
%
&& +  \frac{\sqrt{2}}{a^{2}} \chi'_\nv U_{1,\nv}U_{2,\nv+\xh}
{\cal Q}(U_{1,\nv+\yh}^\dag U_{2,\nv}^\dag
- U_{2,\nv+\xh}^\dag U_{1,\nv}^\dag)
\nn \\
 & & 
+\frac{\sqrt{2}}{a^{2}} \chi'_\nv {\cal Q} 
(U_{1,\nv}U_{2,\nv+\xh} -U_{2,\nv}U_{1,\nv+\yh})
U_{2,\nv+ \xh}^\dag U_{1,\nv}^\dag
%
%
\eea
and
\bea
&& 
-2\sqrt{2}  \chi'_\nv  (\mybar z_{3,\nv}  \chi'_\nv 
-\chi'_\nv U_{1,\nv}^\dag U_{2,\nv+\xh} \mybar z_{3,\nv+\xh+\yh}
U_{2,\nv+\xh}^\dag U_{1,\nv}^\dag)
%
%
\eea
are different from their corresponding terms
$\eta(x) [\phi(x), \eta(x)]$ and
$-i \chi(x) Q \Phi(x)$ in the original Sugino model~(\ref{lat_N=2_S2}). 
In the ${\cal Q}$-transformation laws eq.~(\ref{Q_lattice02}),
only the transformation laws of auxiliary fields 
and its partner $\chi'_\nv$
are different from
ones of the Sugino's model~(\ref{Q_lattice}).
Then the ${\cal N} = (2,2)$ lattice gauge theory of the Sugino type is
derived from the ${\cal N} = (4,4)$ CKKU lattice theory by the 
suitable truncation of fields.

Although we have derived from the CKKU model without moduli fixing
mass term,
we can derive the same model even if 
we introduce the moduli fixing mass term 
\be
\sum_{\nv} \Tr 
\left[\left( z_{i,\nv} \mybar z_{i,\nv} - \frac{1}{2a^2} \right)^2\right]
= \sum_{\nv}\Tr \frac{1}{4}
\left[\left( (s_{i,\nv} + \frac{1}{a})^2 - 
\frac{1}{a^2} \right)^2\right]
\ee
in the CKKU model.
This is because such mass term naturally vanishes under the truncation
$s_{i,\nv} = 0$.

\bibliographystyle{JHEP}

\begin{thebibliography}{10}

\bibitem{Catterall:2004np}
S.~Catterall, {\it A geometrical approach to n = 2 super yang-mills theory on
  the two dimensional lattice},  {\em JHEP} {\bf 11} (2004) 006,
  [\href{http://xxx.lanl.gov/abs/hep-lat/0410052}{{\tt hep-lat/0410052}}].

\bibitem{Sugino:2003yb}
F.~Sugino, {\it A lattice formulation of super yang-mills theories with exact
  supersymmetry},  {\em JHEP} {\bf 01} (2004) 015,
  [\href{http://xxx.lanl.gov/abs/hep-lat/0311021}{{\tt hep-lat/0311021}}].

\bibitem{Cohen:2003qw}
A.~G. Cohen, D.~B. Kaplan, E.~Katz, and M.~Unsal, {\it Supersymmetry on a
  euclidean spacetime lattice. ii: Target theories with eight supercharges},
  {\em JHEP} {\bf 12} (2003) 031,
  [\href{http://xxx.lanl.gov/abs/hep-lat/0307012}{{\tt hep-lat/0307012}}].

\bibitem{Kaplan:2002wv}
D.~B. Kaplan, E.~Katz, and M.~Unsal, {\it Supersymmetry on a spatial lattice},
  {\em JHEP} {\bf 05} (2003) 037,
  [\href{http://xxx.lanl.gov/abs/hep-lat/0206019}{{\tt hep-lat/0206019}}].

\bibitem{Cohen:2003xe}
A.~G. Cohen, D.~B. Kaplan, E.~Katz, and M.~Unsal, {\it Supersymmetry on a
  euclidean spacetime lattice. i: A target theory with four supercharges},
  {\em JHEP} {\bf 08} (2003) 024,
  [\href{http://xxx.lanl.gov/abs/hep-lat/0302017}{{\tt hep-lat/0302017}}].

\bibitem{Kaplan:2005t}
D.~B. Kaplan and M.~Unsal, {\it A euclidean lattice construction of
  supersymmetric yang- mills theories with sixteen supercharges},  {\em JHEP}
  {\bf 09} (2005) 042, [\href{http://xxx.lanl.gov/abs/hep-lat/0503039}{{\tt
  hep-lat/0503039}}].

\bibitem{Endres:2006ic}
M.~G. Endres and D.~B. Kaplan, {\it Lattice formulation of (2,2) supersymmetric
  gauge theories with matter fields},  {\em JHEP} {\bf 10} (2006) 076,
  [\href{http://xxx.lanl.gov/abs/hep-lat/0604012}{{\tt hep-lat/0604012}}].

\bibitem{Damgaard:2007be}
P.~H. Damgaard and S.~Matsuura, {\it Classification of supersymmetric lattice
  gauge theories by orbifolding},
  \href{http://xxx.lanl.gov/abs/arXiv:0704.2696 [hep-lat]}{{\tt arXiv:0704.2696
  [hep-lat]}}.

\bibitem{Sugino:2004qd}
F.~Sugino, {\it Super yang-mills theories on the two-dimensional lattice with
  exact supersymmetry},  {\em JHEP} {\bf 03} (2004) 067,
  [\href{http://xxx.lanl.gov/abs/hep-lat/0401017}{{\tt hep-lat/0401017}}].

\bibitem{Sugino:2004gz}
F.~Sugino, {\it A lattice formulation of super yang-mills theories with exact
  supersymmetry},  {\em Nucl. Phys. Proc. Suppl.} {\bf 140} (2005) 763--765,
  [\href{http://xxx.lanl.gov/abs/hep-lat/0409036}{{\tt hep-lat/0409036}}].

\bibitem{Sugino:2004uv}
F.~Sugino, {\it Various super yang-mills theories with exact supersymmetry on
  the lattice},  {\em JHEP} {\bf 01} (2005) 016,
  [\href{http://xxx.lanl.gov/abs/hep-lat/0410035}{{\tt hep-lat/0410035}}].

\bibitem{Sugino:2006uf}
F.~Sugino, {\it Two-dimensional compact n = (2,2) lattice super yang-mills
  theory with exact supersymmetry},  {\em Phys. Lett.} {\bf B635} (2006)
  218--224, [\href{http://xxx.lanl.gov/abs/hep-lat/0601024}{{\tt
  hep-lat/0601024}}].

\bibitem{Catterall:2005fd}
S.~Catterall, {\it Lattice formulation of n = 4 super yang-mills theory},  {\em
  JHEP} {\bf 06} (2005) 027,
  [\href{http://xxx.lanl.gov/abs/hep-lat/0503036}{{\tt hep-lat/0503036}}].

\bibitem{Catterall:2006jw}
S.~Catterall, {\it Simulations of n = 2 super yang-mills theory in two
  dimensions},  {\em JHEP} {\bf 03} (2006) 032,
  [\href{http://xxx.lanl.gov/abs/hep-lat/0602004}{{\tt hep-lat/0602004}}].

\bibitem{Catterall:2001fr}
S.~Catterall and S.~Karamov, {\it Exact lattice supersymmetry: the
  two-dimensional n = 2 wess-zumino model},  {\em Phys. Rev.} {\bf D65} (2002)
  094501, [\href{http://xxx.lanl.gov/abs/hep-lat/0108024}{{\tt
  hep-lat/0108024}}].
S.~Catterall, {\it Lattice supersymmetry and topological field theory},  {\em
  JHEP} {\bf 05} (2003) 038,
  [\href{http://xxx.lanl.gov/abs/hep-lat/0301028}{{\tt hep-lat/0301028}}].

\bibitem{Unsal:2004cf}
M.~Unsal, {\it Regularization of non-commutative sym by orbifolds with discrete
  torsion and sl(2,z) duality},  {\em JHEP} {\bf 12} (2005) 033,
  [\href{http://xxx.lanl.gov/abs/hep-th/0409106}{{\tt hep-th/0409106}}].
M.~Unsal, {\it Supersymmetric deformations of type iib matrix model as matrix
  regularization of n = 4 sym},  {\em JHEP} {\bf 04} (2006) 002,
  [\href{http://xxx.lanl.gov/abs/hep-th/0510004}{{\tt hep-th/0510004}}].


\bibitem{Giedt:2003ve}
J.~Giedt, {\it Non-positive fermion determinants in lattice supersymmetry},
  {\em Nucl. Phys.} {\bf B668} (2003) 138--150,
  [\href{http://xxx.lanl.gov/abs/hep-lat/0304006}{{\tt hep-lat/0304006}}].
J.~Giedt, {\it The fermion determinant in (4,4) 2d lattice super-yang- mills},
  {\em Nucl. Phys.} {\bf B674} (2003) 259--270,
  [\href{http://xxx.lanl.gov/abs/hep-lat/0307024}{{\tt hep-lat/0307024}}].
J.~Giedt, {\it Deconstruction, 2d lattice yang-mills, and the dynamical lattice
  spacing},  \href{http://xxx.lanl.gov/abs/hep-lat/0312020}{{\tt
  hep-lat/0312020}}.
J.~Giedt, {\it Deconstruction, 2d lattice super-yang-mills, and the dynamical
  lattice spacing},  \href{http://xxx.lanl.gov/abs/hep-lat/0405021}{{\tt
  hep-lat/0405021}}.
J.~Giedt, {\it Deconstruction and other approaches to supersymmetric lattice
  field theories},  {\em Int. J. Mod. Phys.} {\bf A21} (2006) 3039--3094,
  [\href{http://xxx.lanl.gov/abs/hep-lat/0602007}{{\tt hep-lat/0602007}}].
J.~Giedt, {\it Quiver lattice supersymmetric matter, d1/d5 branes and
  ads(3)/cft(2)},  \href{http://xxx.lanl.gov/abs/hep-lat/0605004}{{\tt
  hep-lat/0605004}}.

\bibitem{Suzuki:2005dx}
H.~Suzuki and Y.~Taniguchi, {\it Two-dimensional n = (2,2) super yang-mills
  theory on the lattice via dimensional reduction},  {\em JHEP} {\bf 10} (2005)
  082, [\href{http://xxx.lanl.gov/abs/hep-lat/0507019}{{\tt hep-lat/0507019}}].

\bibitem{Feo:2002yi}
A.~Feo, {\it Supersymmetry on the lattice},  {\em Nucl. Phys. Proc. Suppl.}
  {\bf 119} (2003) 198--209,
  [\href{http://xxx.lanl.gov/abs/hep-lat/0210015}{{\tt hep-lat/0210015}}].
A.~Feo, {\it The supersymmetric ward-takahashi identity in 1-loop lattice
  perturbation theory. i: General procedure},  {\em Phys. Rev.} {\bf D70}
  (2004) 054504, [\href{http://xxx.lanl.gov/abs/hep-lat/0305020}{{\tt
  hep-lat/0305020}}].
A.~Feo, P.~Merlatti, and F.~Sannino, {\it Information on the super yang-mills
  spectrum},  {\em Phys. Rev.} {\bf D70} (2004) 096004,
  [\href{http://xxx.lanl.gov/abs/hep-th/0408214}{{\tt hep-th/0408214}}].
A.~Feo, {\it Predictions and recent results in susy on the lattice},  {\em Mod.
  Phys. Lett.} {\bf A19} (2004) 2387--2402,
  [\href{http://xxx.lanl.gov/abs/hep-lat/0410012}{{\tt hep-lat/0410012}}].

\bibitem{Montvay:2001kw}
I.~Montvay {\em et~al.}, {\it Numerical simulation of supersymmetric yang-mills
  theory}, . Prepared for NIC Symposium 2001, Julich, Germany, 5-6 Dec 2001.
K.~Fujikawa, {\it Supersymmetry on the lattice and the leibniz rule},  {\em
  Nucl. Phys.} {\bf B636} (2002) 80--98,
  [\href{http://xxx.lanl.gov/abs/hep-th/0205095}{{\tt hep-th/0205095}}].
J.~W. Elliott and G.~D. Moore, {\it Three dimensional n = 2 supersymmetry on
  the lattice},  {\em PoS} {\bf LAT2005} (2006) 245,
  [\href{http://xxx.lanl.gov/abs/hep-lat/0509032}{{\tt hep-lat/0509032}}].
H.~Fukaya, I.~Kanamori, H.~Suzuki, M.~Hayakawa, and T.~Takimi, {\it Note on
  massless bosonic states in two-dimensional field theories},  {\em Prog.
  Theor. Phys.} {\bf 116} (2007) 1117--1129,
  [\href{http://xxx.lanl.gov/abs/hep-th/0609049}{{\tt hep-th/0609049}}].
R.~Nakayama and Y.~Okada, {\it Supercurrent anomaly in lattice gauge theory},
  {\em Phys. Lett.} {\bf B134} (1984) 241.
I.~Ichinose, {\it Supersymmetric lattice gauge theory},  {\em Phys. Lett.} {\bf
  B122} (1983) 68.
{\bf DESY-Munster-Roma} Collaboration, F.~Farchioni {\em et~al.}, {\it The
  supersymmetric ward identities on the lattice},  {\em Eur. Phys. J.} {\bf
  C23} (2002) 719--734, [\href{http://xxx.lanl.gov/abs/hep-lat/0111008}{{\tt
  hep-lat/0111008}}].
\bibitem{Kaplan:1999jn}
D.~B. Kaplan and M.~Schmaltz, {\it Supersymmetric yang-mills theories from
  domain wall fermions},  {\em Chin. J. Phys.} {\bf 38} (2000) 543--550,
  [\href{http://xxx.lanl.gov/abs/hep-lat/0002030}{{\tt hep-lat/0002030}}].
Y.~Kikukawa and Y.~Nakayama, {\it Nicolai mapping vs. exact chiral symmetry on
  the lattice},  {\em Phys. Rev.} {\bf D66} (2002) 094508,
  [\href{http://xxx.lanl.gov/abs/hep-lat/0207013}{{\tt hep-lat/0207013}}].
K.~Itoh, M.~Kato, H.~Sawanaka, H.~So, and N.~Ukita, {\it Towards the super
  yang-mills theory on the lattice},  {\em Prog. Theor. Phys.} {\bf 108} (2002)
  363--374, [\href{http://xxx.lanl.gov/abs/hep-lat/0112052}{{\tt
  hep-lat/0112052}}].
G.~T. Fleming, J.~B. Kogut, and P.~M. Vranas, {\it Super yang-mills on the
  lattice with domain wall fermions},  {\em Phys. Rev.} {\bf D64} (2001)
  034510, [\href{http://xxx.lanl.gov/abs/hep-lat/0008009}{{\tt
  hep-lat/0008009}}].
M.~Harada and S.~Pinsky, {\it N = (1,1) super yang-mills on a (2+1) dimensional
  transverse lattice with one exact supersymmetry},  {\em Phys. Lett.} {\bf
  B567} (2003) 277--287, [\href{http://xxx.lanl.gov/abs/hep-lat/0303027}{{\tt
  hep-lat/0303027}}].
M.~Harada and S.~Pinsky, {\it N = 1 super yang-mills on a (3+1) dimensional
  transverse lattice with one exact supersymmetry},  {\em Phys. Rev.} {\bf D71}
  (2005) 065013, [\href{http://xxx.lanl.gov/abs/hep-lat/0411024}{{\tt
  hep-lat/0411024}}].
\bibitem{Montvay:2001aj}
I.~Montvay, {\it Supersymmetric yang-mills theory on the lattice},  {\em Int.
  J. Mod. Phys.} {\bf A17} (2002) 2377--2412,
  [\href{http://xxx.lanl.gov/abs/hep-lat/0112007}{{\tt hep-lat/0112007}}].
Y.~Taniguchi, {\it One loop calculation of susy ward-takahashi identity on
  lattice with wilson fermion},  {\em Chin. J. Phys.} {\bf 38} (2000) 655--662,
  [\href{http://xxx.lanl.gov/abs/hep-lat/9906026}{{\tt hep-lat/9906026}}].
M.~Kato, M.~Sakamoto, and H.~So, {\it Leibniz rule and exact supersymmetry on
  lattice: A case of supersymmetrical quantum mechanics},  {\em PoS} {\bf
  LAT2005} (2006) 274, [\href{http://xxx.lanl.gov/abs/hep-lat/0509149}{{\tt
  hep-lat/0509149}}].
K.~Itoh, M.~Kato, H.~Sawanaka, H.~So, and N.~Ukita, {\it Novel approach to
  super yang-mills theory on lattice: Exact fermionic symmetry and 'ichimatsu'
  pattern},  {\em JHEP} {\bf 02} (2003) 033,
  [\href{http://xxx.lanl.gov/abs/hep-lat/0210049}{{\tt hep-lat/0210049}}].
J.~Nishimura, S.-J. Rey, and F.~Sugino, {\it Supersymmetry on the
  noncommutative lattice},  {\em JHEP} {\bf 02} (2003) 032,
  [\href{http://xxx.lanl.gov/abs/hep-lat/0301025}{{\tt hep-lat/0301025}}].
J.~Nishimura, {\it Four-dimensional n = 1 supersymmetric yang-mills theory on
  the lattice without fine-tuning},  {\em Phys. Lett.} {\bf B406} (1997)
  215--218, [\href{http://xxx.lanl.gov/abs/hep-lat/9701013}{{\tt
  hep-lat/9701013}}].
N.~Maru and J.~Nishimura, {\it Lattice formulation of supersymmetric yang-mills
  theories without fine-tuning},  {\em Int. J. Mod. Phys.} {\bf A13} (1998)
  2841--2856, [\href{http://xxx.lanl.gov/abs/hep-th/9705152}{{\tt
  hep-th/9705152}}].

\bibitem{DAdda:2005zk}
A.~D'Adda, I.~Kanamori, N.~Kawamoto, and K.~Nagata, {\it Exact extended
  supersymmetry on a lattice: Twisted n = 2 super yang-mills in two
  dimensions},  {\em Phys. Lett.} {\bf B633} (2006) 645--652,
  [\href{http://xxx.lanl.gov/abs/hep-lat/0507029}{{\tt hep-lat/0507029}}].
F.~Bruckmann, S.~Catterall, and M.~de~Kok, {\it A critique of the link approach
  to exact lattice supersymmetry},  {\em Phys. Rev.} {\bf D75} (2007) 045016,
  [\href{http://xxx.lanl.gov/abs/hep-lat/0611001}{{\tt hep-lat/0611001}}].

\bibitem{Witten:1988ze}
E.~Witten, {\it Topological quantum field theory},  {\em Commun. Math. Phys.}
  {\bf 117} (1988) 353.

\bibitem{Unsal:2006qp}
M.~Unsal, {\it Twisted supersymmetric gauge theories and orbifold lattices},
  {\em JHEP} {\bf 10} (2006) 089,
  [\href{http://xxx.lanl.gov/abs/hep-th/0603046}{{\tt hep-th/0603046}}].

\bibitem{Douglas:1996sw}
M.~R. Douglas and G.~W. Moore, {\it D-branes, quivers, and ale instantons},
  \href{http://xxx.lanl.gov/abs/hep-th/9603167}{{\tt hep-th/9603167}}.
S.~Kachru and E.~Silverstein, {\it 4d conformal theories and strings on
  orbifolds},  {\em Phys. Rev. Lett.} {\bf 80} (1998) 4855--4858,
  [\href{http://xxx.lanl.gov/abs/hep-th/9802183}{{\tt hep-th/9802183}}].

\bibitem{Arkani-Hamed:2001ie}
N.~Arkani-Hamed, A.~G. Cohen, D.~B. Kaplan, A.~Karch, and L.~Motl, {\it
  Deconstructing (2,0) and little string theories},  {\em JHEP} {\bf 01} (2003)
  083, [\href{http://xxx.lanl.gov/abs/hep-th/0110146}{{\tt hep-th/0110146}}].
N.~Arkani-Hamed, A.~G. Cohen, and H.~Georgi, {\it (de)constructing dimensions},
   {\em Phys. Rev. Lett.} {\bf 86} (2001) 4757--4761,
  [\href{http://xxx.lanl.gov/abs/hep-th/0104005}{{\tt hep-th/0104005}}].

\bibitem{Aratyn:1984bd}
H.~Aratyn, M.~Goto, and A.~H. Zimerman, {\it A lattice gauge theory for fields
  in the adjoint representation},  {\em Nuovo Cim.} {\bf A84} (1984) 255.

\bibitem{Wess-Bagger}
J.~Wess and J.~Bagger, {\it Supersymmetry and supergravity}, .

\bibitem{Unsal:2005yh}
M.~Unsal, {\it Compact gauge fields for supersymmetric lattices},  {\em JHEP}
  {\bf 11} (2005) 013, [\href{http://xxx.lanl.gov/abs/hep-lat/0504016}{{\tt
  hep-lat/0504016}}].

\bibitem{Onogi:2005cz}
T.~Onogi and T.~Takimi, {\it Perturbative study of the supersymmetric lattice
  theory from matrix model},  {\em Phys. Rev.} {\bf D72} (2005) 074504,
  [\href{http://xxx.lanl.gov/abs/hep-lat/0506014}{{\tt hep-lat/0506014}}].

\bibitem{Ohta:2006qz}
K.~Ohta and T.~Takimi, {\it Lattice formulation of two dimensional topological
  field theory},  {\em Prog. Theor. Phys.} {\bf 117} (2007) 317--345,
  [\href{http://xxx.lanl.gov/abs/hep-lat/0611011}{{\tt hep-lat/0611011}}].

\end{thebibliography}


\providecommand{\href}[2]{#2}\begingroup\raggedright\endgroup

\end{document}